%% file: summary.tex
\documentclass[final]{IEEEtran}
\usepackage{amsthm,amssymb,graphicx,multirow,amsmath,color,amsfonts}
\usepackage[update,prepend]{epstopdf}
\usepackage[noadjust]{cite}
\usepackage[latin1]{inputenc}
\usepackage{tikz}
\usepackage{bbm} 
\usepackage{pdfpages}
\usepackage{tabulary}
\usepackage{multirow}
\usepackage{comment}
\usepackage{subfigure}
\usepackage{float}

\include{notation}

\allowdisplaybreaks 
\begin{document}
	\title{Performance  Analysis of Charging Infrastructure Sharing in UAV and EV-involved Networks}
		\author{
		Yujie Qin, Mustafa A. Kishk, {\em Member, IEEE}, and Mohamed-Slim Alouini, {\em Fellow, IEEE}
		\thanks{Yujie Qin and Mohamed-Slim Alouini are with Computer, Electrical and Mathematical Sciences and Engineering (CEMSE) Division, King Abdullah University of Science and Technology (KAUST), Thuwal, 23955-6900, Saudi Arabia
			Arabia. Mustafa Kishk was with CEMSE, KAUST during this work. Currently  he is with the Department of Electronic Engineering, National University of Ireland, Maynooth, W23 F2H6, Ireland. (e-mail: yujie.qin@kaust.edu.sa; mustafa.kishk@mu.ie; slim.alouini@kaust.edu.sa).} 
		
	}
	\date{\today}
	\maketitle
	
\begin{abstract}
    	Electric vehicles (EVs) and unmanned aerial vehicles (UAVs) show great potential in modern transportation and communication networks, respectively. However, with growing demands for such technologies, the limited energy infrastructure becomes the bottleneck for their future growth. It might be of high cost and low energy efficiency for all the operators to each have their own dedicated energy infrastructure, such as charging stations. In this paper, we analyze a wireless charging infrastructure sharing strategy in UAV and EV-involved networks. We consider a scenario where UAVs can charge in EV charging stations and pay for the sharing fee. On the EVs' side, sharing infrastructure can earn extra profit but their service quality, such as waiting time, might slightly reduce. On the UAVs' side, if renting EV charging stations can achieve an acceptable system performance, say high coverage probability, while considering the cost, they may not need to build their dedicated charging stations. In this case, we use tools from stochastic geometry to model the locations and propose an optimization problem that captures the aforementioned trade-offs between cost or profit and quality of service. Our numerical results show that sharing infrastructure slightly increases the waiting time of EVs, say within $5$ min,  but dramatically decreases the waiting time of drones, say more than $50$ min, and deploying more charging stations do achieve better performances, but all these better performances are expected to cost more. 
\end{abstract}
	
	
	\section{Introduction}
Electric vehicles (EVs) and unmanned aerial vehicles (UAVs) have great application potential to achieve green and efficient future transportation \cite{gan2012optimal}. 	Compared with traditional vehicles, EVs cause less impact on the environment and are more flexible since they can recharge at home during the night. Hence, in the past few decades, EVs and their related infrastructures, such as charging stations, have been widely studied and developed in real life \cite{liu2021reservation}. 

UAVs have recently increased in the market due to their high relocation flexibility based on dynamic demands\cite{sekander2018multi,mozaffari2019tutorial}. They are expected to play an essential rule in next-generation wireless networks \cite{li2018uav,zeng2016wireless}. UAVs can be very useful in both rural and urban areas \cite{9773146}, such as improving the quality of service \cite{liao2020learning}, providing service to remote Internet of Things users \cite{cheng2019space}, maintaining connectivity in nature disasters \cite{matracia2021topological} and providing additional capacity \cite{9205314}. However, UAVs cannot work without frequent charging, which highly restricts the performance of UAV-enabled wireless networks. Since UAVs rely on their internal batteries, which are energy limited, their services are likely to be interrupted if the energy runs low. Meanwhile, compared with EVs, UAV-assisting networks are still new, and their infrastructures are poor. While researchers mainly focus on investigating the communication or application parts of UAV-involved networks, the infrastructures of UAVs, such as the deployment of charging stations, should catch up \cite{9773146}.

The traditional model of single ownership of all the physical infrastructures now faces significant challenges: it may be high cost and waste lots of energy. Take charging stations for instance, the cost of building new charging stations includes installation, maintenance, electricity grid distribution, storage \cite{8673613}, resulting in huge pressure on operators.  Meanwhile, it is not energy-efficient for all operators to each have their own grid infrastructure. In the current market environment, infrastructure sharing idea may be a sustainable way to follow. Instead of building dedicated charging stations, operators can consider sharing the charging stations while maintaining an acceptable service quality.

In this paper, we explore an infrastructure sharing strategy in the EVs and UAVs-involved network: EVs' operators are willing to share their charging stations with UAVs as far as their services are acceptable, and UAVs pay for the sharing. In other words, UAVs can charge in EV charging stations to avoid the traffic, hence, to achieve a better system performance. However, the charging time may be longer in EV charging stations since they are not designed for UAVs. We would like to emphasize that we are considering wireless charging in this work which enable using the charging station by two different technologies such as the UAVs and the EVs. More details are presented later in this section and Section \ref{sec_inf_sharing}.

	\subsection{Related Work}
Literature related to this work can be categorized into (i) design of EV charging stations, (ii) UAV-assisted networks, and (iii) design of infrastructure sharing. We provide a brief introduction to each of them in the following lines.

{\em EV charging stations placement and charging schedule.} A comprehensive survey on EV transportation was provided in \cite{SHAUKAT20181329}, which was mainly about key technologies, such as energy storage, for transportation electrification in smart grid scenarios. EV charging infrastructure deployment was analyzed in \cite{sweda2011agent}. Optimization of waiting time for EVs is studied in \cite{9681355} using a fuzzy inference system. A Japanese case study about quick charging of EVs based on waiting time and cost-benefit analyses is investigated in \cite{oda2018mitigation}. The authors studied the infrastructure placement from the perspective of the agent and used the road network data of the Chicagoland area. Authors in \cite{sathaye2013approach} presented an optimization of minimum plug-in electric vehicles (PEV) infrastructure deployment for highway corridors. Their models allowed for a straightforward analysis despite uncertain input data such as the uncertainty of unavailable information on PEV drivers' behavior and charging demand data. Besides, much of existing work on EVs is related to charging scheduling in a vehicle-to-grid (V2G) system, such as \cite{6486056,6557596}. The authors used queuing networks to model the dynamics of EVs and investigated the joint scheduling, which allowed the operators to optimize the total cost of the system \cite{ 6486024}. 

{\em Stochastic geometry-based UAV-enabled network analysis.} Stochastic geometry is a strong mathematical tool that enables large-scale wireless networks and has been demonstrated that it provides a tight approximation to real networks. In-depth tutorial and survey about modeling base stations (BSs) and characterizing interference had been provided in \cite{7733098,6524460}. Authors in \cite{9153823,9444343} modeled the locations of UAVs and charging stations by two independent Poisson point processes (PPPs) and modified the definition of coverage probability based on queuing theory by considering the energy limitation of UAVs. In \cite{8866716}, the authors proposed a laser-powered UAV system and introduced a new concept of energy coverage probability, which is a joint probability of harvesting energy and SNR coverage. Authors in \cite{sekander2020statistical} considered renewable energy-powered UAVs, which can harvest energy from solar or wind resources, and derived the probability density function (PDF) and cumulative density function (CDF) of harvest energy and outage probability. Besides harvesting energy, authors in \cite{kishk20203,9205314,lou2021green} studied the tethered UAV, which is physically connected to a ground station. While the tether provides the UAV with a stable power supply and reliable data rate, it highly restricts the mobility and freedom of UAVs.

{\em Economic analysis of wireless network infrastructure sharing.} A brief review about infrastructure sharing was provided in \cite{meddour2011role}, which captured a conflict between high demands of infrastructures and high cost. They categorized four types of infrastructure sharing models, analyzed them from the perspective of economic dimensions, and provided a practical use case. Authors in \cite{sanguanpuak2018infrastructure} modeled and analyzed an infrastructure sharing system composed of a single buyer mobile network operator and multiple sellers. Specifically, they modeled the locations of BSs by PPPs and found the coverage probability of downlink signal-to-interference-plus-noise (SINR) under the sharing environment. Using game theory, authors in \cite{bousia2015game} proposed a system with a switching off decision, which enabled the operators to switch off some BSs during low traffic time, such as night. Besides BSs, spectrum licenses can also be shared as mentioned in \cite{8315130}. Similarly, authors in \cite{fund2016spectrum} studied the spectrum sharing in Millimeter-wave cellular networks from the perspective of economic.

While most of the existing literature considers the energy resources separately, none of them analyze the possibility of energy infrastructure sharing.

	\subsection{Contribution}
In this paper, our main goal is to explore a charging infrastructure sharing strategy among the EV and UAV-involved network, based on operators' decisions, high profit, or good service quality.  We tap a new concept which is charging infrastructure sharing among UAVs and EVs for the first time. Since it is a totally new idea, it is difficult to find supporting data,  and our system models and analysis are based on some reasonable assumptions. The main contributions of this work included the following points.

{\em Modeling of waiting time.} While we consider that UAVs can charge in EV charging stations, the proposed system combines M/G/$c$ and D/D/$c$ queues, waiting times of both EVs and UAVs become complex to model. We derive some tight approximate equations about the waiting time of EVs and UAVs in continuous time cases using the results from renewal processes and under some reasonable assumptions. We then show the  waiting time gap between the approximated analysis and simulations.

{\em Coverage probability.} We consider two association policies between UAVs and charging stations: (i) biased distance, in which the association cells form a multiplicatively weighed Cox-Voronoi tessellation, and (ii) independent thinning, in which the association regions form two independent Cox-Voronoi tessellations. Our results show that association based on independent thinning is slightly worse, however, easier than biased distance. Building upon these two association policies and the waiting time we derived, we formulate a more accurate expression for coverage probability. It captures the queuing of UAVs and the influence of limited energy resources. 

{\em Economic insights based on charging infrastructure sharing.} We consider two scenarios where the operators care more about their service or profit. The established optimization problem analyzes the weights of cost, profit, quality of service from the perspective of EVs and UAVs, respectively. Our results show that charging infrastructure sharing benefits both UAVs and EVs operators, especially when the quality of the charging station is high, say can charge multiple EVs or UAVs.  More details shown in Section \ref{sec_num}.
	
	\section{System Model}
	  	\begin{table*}\caption{Table of Notations}\label{table_par_ana}
	\centering
	\begin{center}
		\resizebox{2\columnwidth}{!}{
			\renewcommand{\arraystretch}{1}
			\begin{tabular}{ {c} | {c} }
				\hline
				\hline
				 \textbf{Notation}&\textbf{Description}  \\ \hline
				$\Phi_{ c,ev}$, $\Phi_{ c,d}$; $\Phi_{ t}$ & Locations of UAV, EV charging stations; TBSs \\ \hline
				 $\lambda_{ l}$; $\lambda_{ p,ev}$, $\lambda_{ p,d}$ & Line density, point density \\ \hline
				 $\lambda_{ c,d}$, $\lambda_{ c,ev}$& UAV, EV charging station density \\ \hline
				$\lambda_{ t}$, $\lambda_{ u}$&  TBS density, UAV density \\ \hline
				 $\mu$, $\sigma$& Average and deviation of SOC \\ \hline
				 $P_{\rm cha}$& EV charging rate \\ \hline
				 $a,b$& CV (or MWCV) tessellation fitting parameters \\ \hline
				 $p_{ m}$, $p_{ s}$& Traveling-related, service-related power \\\hline
				 $h$, $v$& UAV altitude, velocity during traveling \\\hline
				 $B_{\rm max,ev}$, $B_{\rm max}$ &  EV, UAV battery capacity\\\hline
				 $T_{\rm ch,d,ev}$, $T_{\rm ch,d,d}$; $T_{\rm tra}$, $T_{\rm ser}$& Charging time of EVs, UAVs; service, traveling time of UAVs \\\hline
				 $R_{ s,d}$, $R_{ s,ev}$  & Distance to the nearest UAV, EV charging station\\\hline
				 $r_c$& Radius of MCP disk \\\hline
				 $c_1, c_2$ & N/LoS environment variable \\\hline
				 $\rho_{ u}$, $\rho_{ t}$& Transmission power of UAVs, TBSs, respectively \\\hline
				 $\gamma$, $\sigma^2_n $& SINR threshold, noise power \\\hline
				 $\alpha_{ n},\alpha_{ l},\alpha_{ t}$ & N/LoS and active charging station path-loss exponent \\\hline
				  $m_{ n},m_{ l}$, $\eta_{ n},\eta_{ l}$ & N/LoS fading gain, additional loss
				\\\hline
				  $\mathcal{A}_{\rm LoS}(r)$, $\mathcal{A}_{\rm NLoS}(r)$, $\mathcal{A}_{\rm TBS}(r)$ & Probability of the reference user associating with nearest LoS/NLoS UAVs, or TBS, respectively
				\\\hline
			 $\mathcal{A}_{d}(\beta_d)$, $\mathcal{A}_{ev}(\beta_d)$& 	Probability of the reference UAV associating with EV, UAV charging stations 
				\\\hline
				  $N_{d,ev}$, $N_{d,d}$&Number of drones in EV, UAV charging stations 
				\\\hline
				$T_{\rm w,ev}$; $T_{\rm w,d,d}$, $T_{\rm w,d,ev}$&  Waiting time of EVs, waiting time of UAVs in UAV, EV charging stations, respectively 
				\\\hline\hline
		\end{tabular}}
	\end{center}
\end{table*}

\begin{figure}[ht]
	\centering
	\includegraphics[width = \columnwidth]{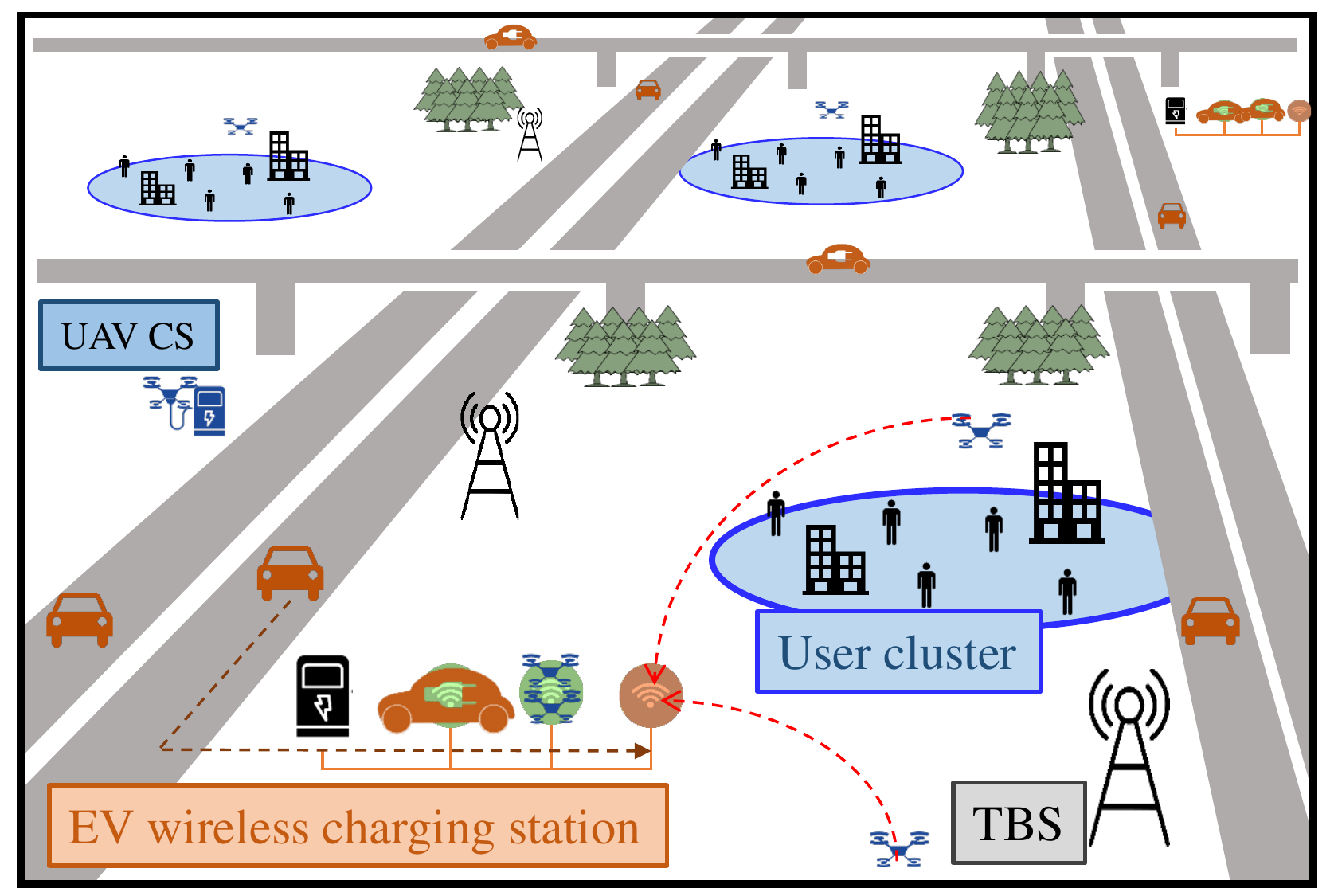}
	\caption{Illustration of the system network.}
	\label{fig_sys_3D}
\end{figure}

We consider a network composed of EVs, UAVs, their own charging stations $\Phi_{ c,ev}$ and $\Phi_{ c,d}$,  and terrestrial base stations (TBSs) $\Phi_{ t}$. The system is shown in Fig. \ref{fig_sys_3D} and related notations are explained in Table \ref{table_par_ana}. The locations of EV charging stations and UAV charging stations are modeled as two independent Poisson line cox processes (PLCPs) $\Phi_{ c,ev}$ and $\Phi_{ c,d}$ with the same line density $\lambda_{ l}$ and different point densities $\lambda_{ p,ev}$ and $\lambda_{ p,d}$. While we consider that EVs are more widely used than UAVs, $\lambda_{ p,ev}$ is higher than $\lambda_{ p,d}$. That is,  $\lambda_{ c,ev}>\lambda_{ c,d}$, where $\lambda_{ c,ev}$ and $\lambda_{ c,d}$ are densities of EV and UAV charging stations, given by $\lambda_{ c} = \pi\lambda_{ l}\lambda_{ p}$.

\begin{figure*}[ht]
	\centering
	\subfigure[]{\includegraphics[width = 1\columnwidth]{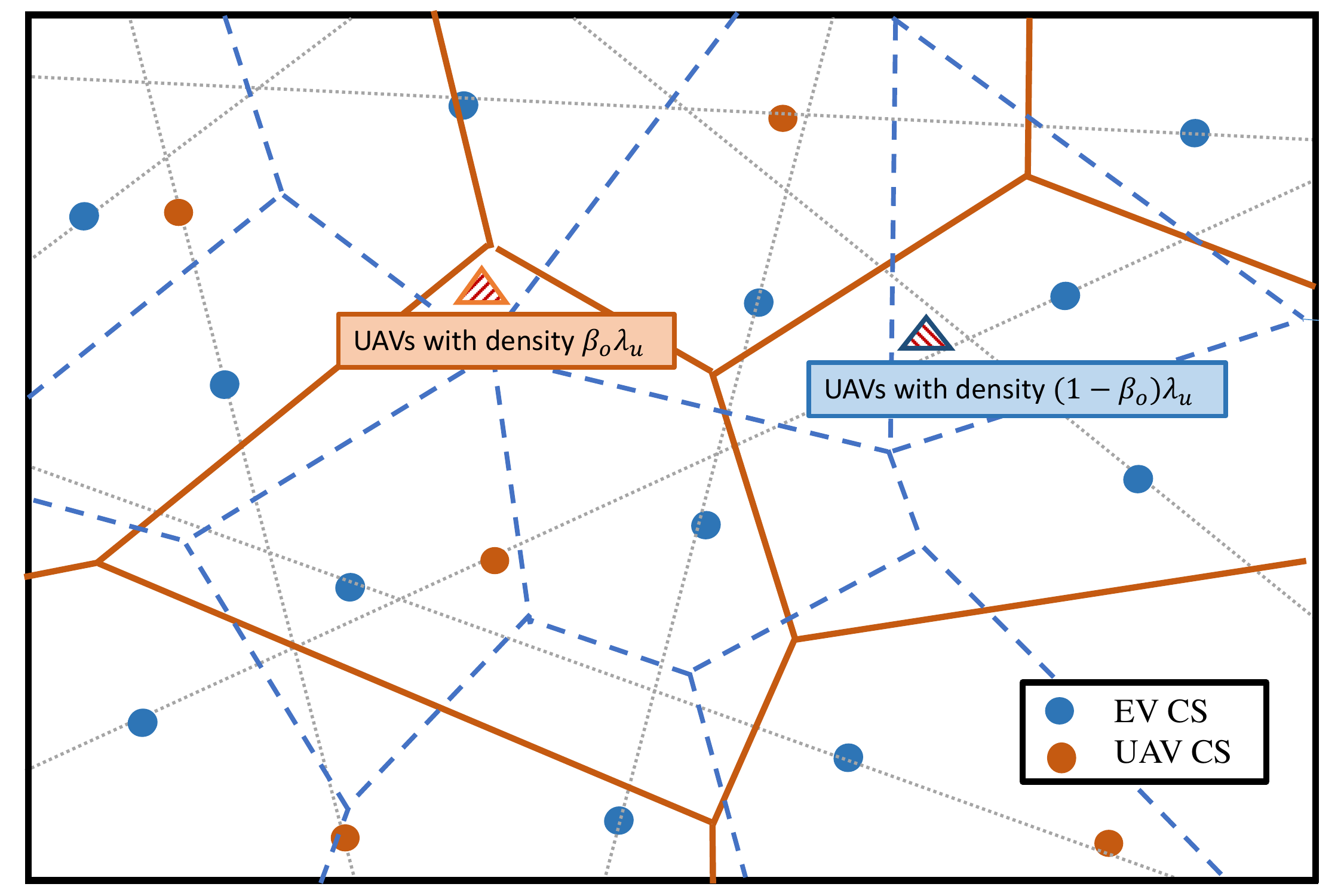}}
	\subfigure[]{\includegraphics[width = 1\columnwidth]{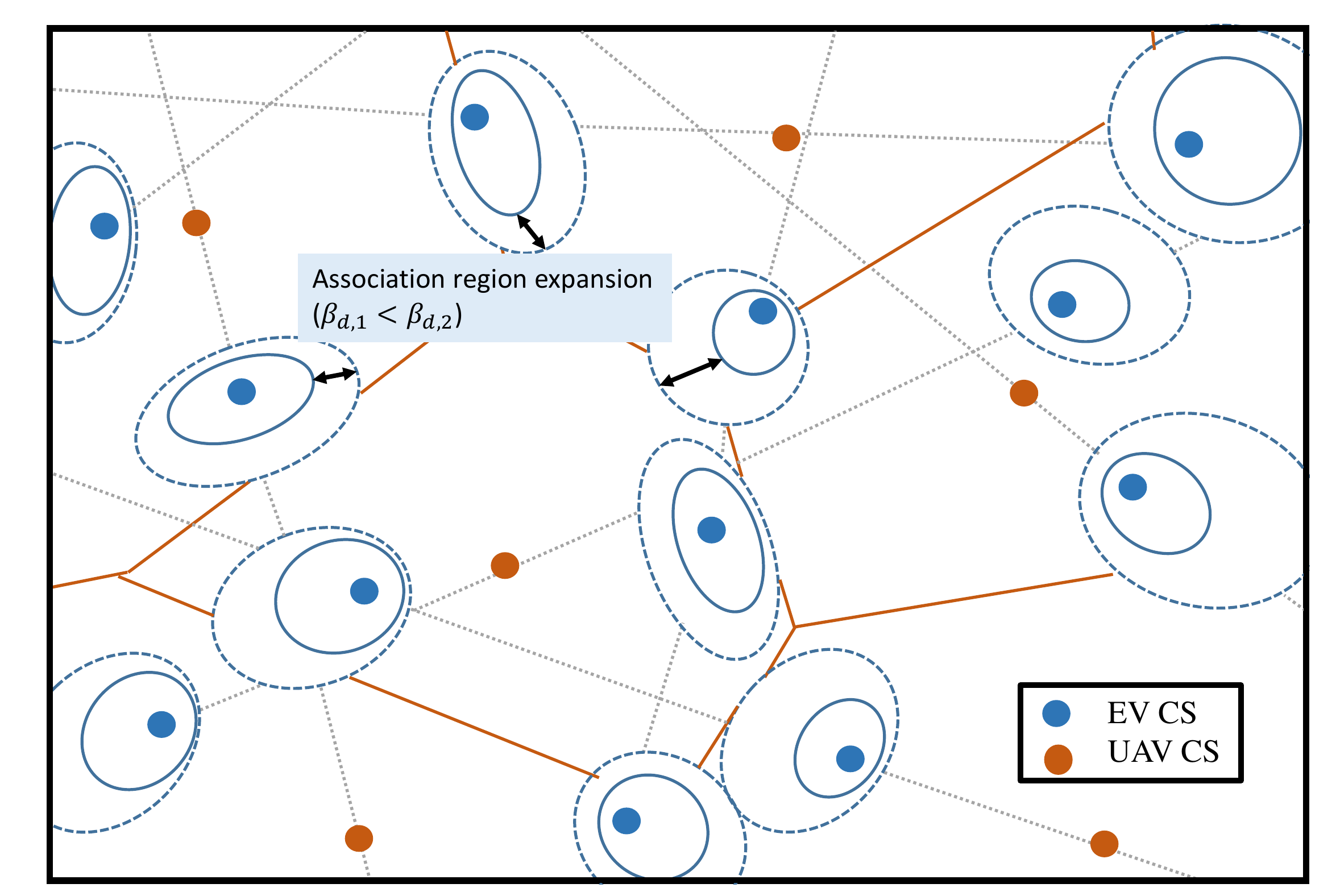}}
	\caption{Illustration of the system model, two types of UAVs' association policies. \textbf{(a)} association based on independent thinning: the association cell forms two independent CV tessellations. \textbf{(b)} association based on biased distance: the association cell forms one MWCV tessellation. }	\label{sys_mod2}
\end{figure*}

The locations of UAVs are modeled as a PPP $\Phi_{ u}$ with density $\lambda_{ u}$. Given that UAVs are hovering at a fixed altitude $h$ above user cluster centers to provide service and they only travel to the charging station to recharge/swap their battery. 
To analyze the benefits of infrastructure sharing, we assume that UAVs can recharge in both types of charging stations and we analyze two different association policies, as shown in Fig. \ref{sys_mod2}: (i) independent thinning, in which a fraction of UAVs recharge in EV charging stations denoted by the offloading ratio $\beta_o$, and (ii) biased distance, in which UAVs associate with the charging station based on the biased distance, $\min(R_{ s,ev},\beta_d R_{ s,d})$, where $R_{ s,ev}$ and $R_{ s,d}$ are the distances between the UAV and its nearest EV/UAV charging station, respectively, and $\beta_d$ is the association weight. In the first policy,  the densities of UAVs recharge in EV charging stations and UAV charging stations are $\beta_o\lambda_{ u}$ and $(1-\beta_o)\lambda_{ u}$, respectively, and the association regions of the UAVs with two types of charging stations form two independent Cox-Voronoi (CV) tessellations. In the second policy, the association regions form one multiplicatively weighed Cox-Voronoi (MWCV) tessellation. Notice that $\beta_d = 1$ is the special case of the presented model with equal distance: UAVs go to the nearest charging stations. Without loss of generality, we perform our analysis in the rest of the paper at a typical UAV located at the origin and the typical association (CV or MWCV) region that contains the origin, denoted by association cell.

The arrival process of EVs is considered as a Poisson process with an average arrival rate $\mu_{ e}$ and the EV charging stations charge EVs and UAVs depending on different serving policies. Moreover, we assume that EV charging stations can serve $c$ customers simultaneously, which is also known as the  capacity of EV charging stations, and $m$ UAVs can be charged together, due to the size of EVs being larger than UAVs. It means that while we assume that UAVs can recharge in EV charging stations, say two or three UAVs can charge within the same time slot. To be more realistic, our system also includes TBSs, to serve the users in the clusters, whose locations are modeled by an independent PPP with density $\lambda_{ t}$.
%
	
	\subsection{Waiting Time}
	
	The remaining state of charging (SOC) of the EV battery level is a random variable modeled by a truncated lognormal distribution with average ($\mu$) and typical deviation ($\sigma$) \cite{dominguez2019design} as follows
	\begin{align}
		f_{\rm SOC}(e) = \frac{1}{e\sigma\sqrt{2\pi}}\frac{\exp(-\frac{(\ln e-\mu)^2}{2\sigma^2})}{F_{\rm SOC}(100)-F_{\rm SOC}(0)},\label{eq_fsoc}
	\end{align}
where $F_{\rm SOC}(e) = \frac{1}{2}\bigg[1+\erf\bigg(\frac{\ln e-\mu}{2\sigma^2}\bigg)\bigg]$.
	Assume that the EVs are fully charged when they leave the charging station, and the charging time $T_{\rm ch,ev}$ is
	\begin{align}
		T_{\rm ch,ev} = \frac{B_{\rm max,ev}}{P_{\rm cha}}(1-\frac{{\rm SOC}}{100}),
	\end{align}
	where $P_{\rm cha}$ is the charging rate.
	
\begin{figure}
		\centering
		\includegraphics[width = \columnwidth]{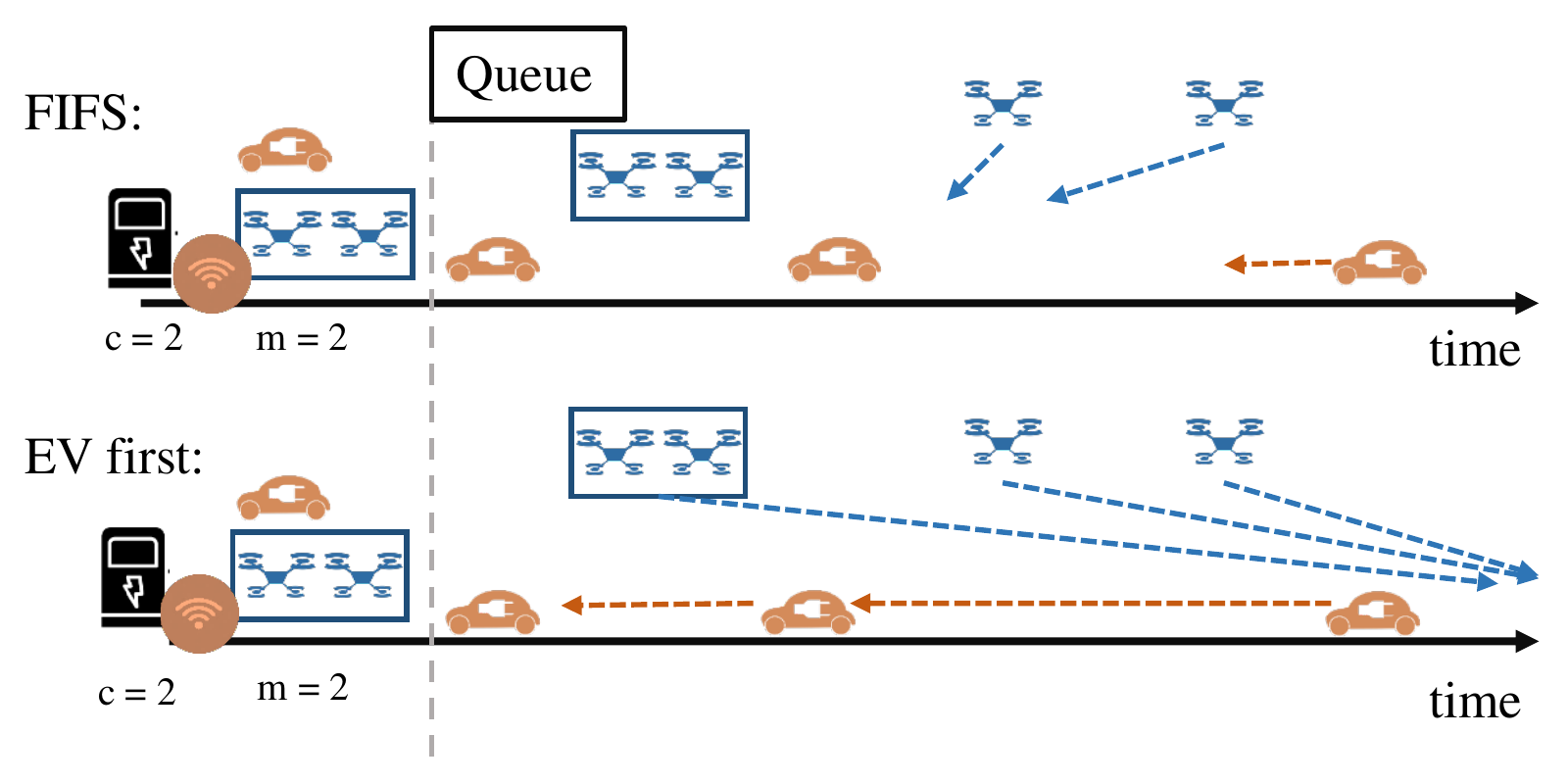}
		\caption{Illustration of the system model, two types of serving policies: (i) charge depends on arrival time and (ii) EVs have higher priority.}
		\label{fig_sys_mod1}
	\end{figure}

	As mentioned, UAVs can recharge in both types of charging stations, $\Phi_{ c,ev}$ and $\Phi_{ c,d}$ with different charging time $T_{\rm ch,d,ev}$ and $T_{\rm ch,d,d}$, respectively. We compare the system performance under two serving policies: first in first serve (FIFS) and EV first, as shown in Fig. \ref{fig_sys_mod1}.
	
Recall that EV charging stations have capacity $c$ and can charge $m$ UAVs simultaneously.
	\begin{definition}[FIFS]\label{def_fifs}
		In the case of FIFS, $\Phi_{ c,ev}$ charge EVs and UAVs based on their arrival time. In this case, the waiting time of UAVs and EVs are respectively denoted by $T_{\rm w,ev,fifs}$ and $T_{\rm w,d,fifs}$.
	\end{definition}
	
	\begin{definition}[EV First]\label{def_evfirst}
		In the case of EV first, EVs have higher priority than UAVs: $\Phi_{ c,ev}$ charge the EVs first and UAVs can only be charged when no EV waiting.
		In this case, the waiting time of UAVs and EVs are respectively denoted by $T_{\rm w,ev,evfirst}$ and $T_{\rm w,d,evfirst}$.
	\end{definition}

		Assume that the UAV is available when it is hovering above the cluster center  and providing service, and is unavailable when traveling to the nearest charging station and waiting to recharge. Hence, the availability probability of a UAV is a time fraction.
	
	\begin{definition}[Availability Probability]
		We define the event $\mathcal{A}$ that indicates the availability of the typical UAV, which denotes that the UAV is available and can provide service. Conditioned on  $N$ UAVs in the typical association cell, the availability probability of the UAV is
		\begin{align}
			\mathbb{P}({ \mathcal{A} | N})= \mathbb{E}_{ \Phi_{c}}\bigg[\frac{T_{\rm ser}(x)}{T_{\rm ser}(x)+T_{\rm ch}+T_{\rm w,d\mid N}+2T_{\rm tra}(x)}\bigg],\label{eq_P_aN}
		\end{align}
		where $x$ annotates the location of the typical UAV, and 
		\begin{align}
			T_{\rm tra}(x)&=\frac{R_{ s,\{ev,d\}}(x)}{v}\label{eq_Ttra},\\
			T_{\rm ser}(x)&=\frac{B_{\rm max}-2p_{ m}\frac{R_{ s,\{ev,d\}}(x)}{v}}{p_{ s}},\label{eq_Tse}
		\end{align}
	in which $B_{\rm max}$, $v$, $p_m$ and $p_s$ is the maximum battery size, velocity, traveling- and serving-related power consumption of UAVs, and waiting time of UAVs depends on $N$, either $T_{\rm w,d,fifs\mid N}$ or $T_{\rm w,d,evfirst\mid N}$, and $N$ can be either $N_{d,ev}$ or $N_{d,d}$ , which are explained in Section \ref{sec_waitingtime}. Hence, the unconditional availability probability is 
		\begin{align}
			P_{\rm a}&=\mathbb{E}_{ N}[\mathbb{P}( \mathcal{A} | N)].\label{def_eq_Pa}
		\end{align}
		
	\end{definition}
    \subsection{Power Consumption}

We consider the UAV's power consumption composed of two parts: (i) service-related power $p_{ s}$, including hovering and communication-related power, (ii) traveling-related power $p_{ m}$, which denotes the power consumed in traveling to/from the   EV/UAV charging station  through the horizontal distance $R_{ s,\{ev,d\}}$.

Given in \cite{zeng2019energy}, ${p}_{ m}$ is a function of the UAV's velocity $v$ and given by
\begin{align}
	p_{ m}=P_{ 0}\left(1+\frac{3v^2}{U_{\rm tip}^{2}}\right)+\frac{P_{ i}v_{ 0}}{v}+\frac{1}{2}d_{ 0}\rho s Av^{3},
\end{align}
where $P_{ 0}$ and $P_{ i}$ present the blade profile power and induced power, $U_{\rm tip}$ is the tip speed of the rotor blade, $v_{ 0}$ is the mean rotor induced velocity in hover, $\rho$ is the air density, $A$ is the rotor disc area, $d_{ 0}$ is fuselage drag ratio, and $s$ is rotor solidity. In this case, the energy consumed during traveling to or from the associated charging station is 
\begin{align}
	E_t&=\frac{R_{ s,\{ev,d\}}(x)}{v}p_{ m} \nonumber\\
	&=\frac{R_{ s,\{ev,d\}}(x)}{v}\left(P_{ 0}\left(1+\frac{3v^2}{U_{\rm tip}^{2}}\right)+\frac{P_{ i}v_{ 0}}{v}+\frac{1}{2}d_{ 0}\rho s Av^{3}\right).
\end{align}
In the rest of the paper, we use the optimal value of $v$ that minimizes $E_t$.

	\subsection{User Association}
	
\label{user_association}
Without loss of generality, we focus on a typical user randomly selected from the typical user cluster centered at the origin. From the perspective of the typical user, we denote the typical UAV as cluster UAV. Assume that the user associates with the cluster UAV if it is available (hovering and providing service), if not, associates with a nearby available UAV or TBS, depending on the average received power.
Let $\Phi_{ u_{o}}$,  $\Phi_{ u^{'}}$ and  $\Phi_{ t}$ be the point sets of serving or interference BSs: cluster UAV, available UAVs and TBSs. Notice that the set $\Phi_{ u_{o}}$ is composed of only one point, which is the location of the cluster UAV in case it is available, otherwise, $\Phi_{ u_{o}}=\emptyset$. When the cluster UAV is unavailable, the user associates with a UAV in $\Phi_{ u^{'}}$ or the closest TBS in $\Phi_{ t}$. The point process $\Phi_{ u^{'}}$ is constructed by independently thinning $\Phi_{ u}$ with the probability $P_{ a}$. Therefore, the density of $\Phi_{ u^{'}}$ is $\lambda_{u}^{'}=P_{ a}\lambda_{ u}$.

When the typical user associates with a UAV, it can be either line-of-sight (LoS) or non line-of-sight (NLoS), the received power is
\begin{align}
	p_{ u}&=\left\{ 
	\begin{aligned}
		p_{ l}=\eta_{ l}\rho_{ u}G_{ l}R_{ u}^{-\alpha_{ l}},  & \quad \text{\rm in case of LoS},\\
		p_{ n}=\eta_{ n}\rho_{ u}G_{ n}R_{ u}^{-\alpha_{ n}},  & \quad \text{\rm in case of NLoS},\\
	\end{aligned} \right.\nonumber
\end{align}
where $\rho_{ u}$ is the transmission power of the UAVs, $R_{ u}$ denotes the Euclidean distance between the typical user and the serving UAV, $\alpha_{ l}$ and $\alpha_{ n}$ present the path-loss exponent, $G_{ l}$ and $G_{ n}$ are the fading gains that follow gamma distribution with shape and scale parameters $(m_{ l},\frac{1}{m_{ l}})$ and $(m_{ n},\frac{1}{m_{ n}})$, $\eta_{ l}$ and $\eta_{ n}$ denote the mean additional losses for LoS and NLoS transmissions, respectively. Based on \cite{al2014optimal}, the probability of establishing LoS or NLoS channels between users and UAVs is
	\begin{align}
		P_{ l}(r) & =  \frac{1}{1+c_1 \exp\bigg(-c_2\bigg(\frac{180}{\pi}\arctan\bigg(\frac{h}{\sqrt{r^2-h^2}}\bigg)-c_1\bigg)\bigg)} ,\label{pl_pn}
	\end{align}
where $r$ is the Euclidean distance, $c_1$ and $c_2$ are two environment-related variables (e.g., urban, dense urban, and highrise urban), and $h$ is the altitude of the UAV. Consequently, the probability of NLoS is $P_{ n}(r)=1-P_{ l}(r)$.

When the user associates with TBS, the received power is
\begin{align}
	p_{ t} &= \rho_{ t} H R_{ t}^{-\alpha_{ t}},\nonumber
\end{align}
in which $\rho_{ t}$ is the transmission power of TBSs, $R_{ t}$ denotes the distance between the user and the nearest TBS, $H$ is the fading gain that follows exponential distribution with unity mean, and $\alpha_{ t}$ presents the path-loss exponent.

Let $\mathcal{A}_{\rm NLoS}(r)$, $\mathcal{A}_{\rm LoS}(r)$ and  $\mathcal{A}_{\rm TBS}(r)$ be the probabilities that the reference user associates with the nearest N/LoS UAV or TBS, which is at $r$ away, are respectively given by
\begin{align}
	\mathcal{A}_{\rm \{NL,L\}oS}(r) &= \mathbb{P}(p_{\{n,l\}}(r)>p_t),\nonumber\\
	\mathcal{A}_{\rm TBS}(r) &= \mathbb{P}(p_t(r)>p_u).
\end{align}

We then define the coverage probability as the probability that the typical user is successfully served, which is the event that SINR of the related link is above a predefined threshold.
\begin{definition}[Coverage Probability] 
	\label{def_cov}
	The total coverage probability is defined as 
\begin{align}
	P_{\rm cov} =& P_{a}P_{\rm cov,U_o}+(1-P_{a})P_{\rm cov,\hat{U}_o}\nonumber\\
	=& P_{a}(P_{\rm cov,U_o,l}+P_{\rm cov,U_o,n})\nonumber\\
	&+(1-P_{a})(P_{\rm cov,\hat{U}_o,l}+P_{\rm cov,\hat{U}_o,n}+P_{\rm cov,t}),\label{eq_Pcov}
\end{align}
	in which,
	\begin{align}
		P_{\rm cov,{U_o}} &= \sum_{bs\in\{\text{\rm N/LoS}\}}\mathbb{E}[\mathcal{A}_{bs}(r)\mathbb{P}({\rm SINR}\geq \gamma|r,bs)],\nonumber\\
		P_{\rm cov,{\hat{U}_o}} &= \sum_{bs\in\{\text{\rm TBS,N/LoS}\}}\mathbb{E}[\mathcal{A}_{bs}(r)\mathbb{P}({\rm SINR}\geq \gamma|r,bs)],\nonumber
	\end{align}
	where $P_{\rm cov,{U_o}}$ and $P_{\rm cov,{\hat{U}_o}}$ are the coverage probabilities when the cluster UAV is available and unavailable, respectively. $P_{\rm cov,U_o,l}$ and $P_{\rm cov,U_o,n}$ are the coverage probabilities when associating with the  LoS/NLoS cluster UAV. $P_{\rm cov,\hat{U}_o,l}$, $P_{\rm cov,\hat{U}_o,n}$ and $P_{\rm cov,t}$ are the coverage probabilities when associating with  the nearby available LoS/NLoS UAV and the nearest TBS, respectively. 
	
	$\Phi_{ u^{'}}$ is composed of two subsets $\Phi_{ u^{'}_l}$ and $\Phi_{ u^{'}_n}$, which denote the locations of available LoS UAVs and NLoS UAVs, respectively.  Conditioning on the serving BS $b_s$, the SINR and the aggregate interference is defined as
	\begin{align}
		{\rm SINR} &= \frac{\max(p_u,p_t)}{I+\sigma^2},\nonumber\\
		I &= \sum_{N_i\in\Phi_{ u^{'}_n}/b_{s}}\eta_{ n}\rho_{ u}G_{ n}D_{ N_i}^{-\alpha_{ n}}+\sum_{L_j\in\Phi_{ u^{'}_l}/b_{s}}\eta_{ l}\rho_{ u}G_{ l}D_{ L_j}^{-\alpha_{ l}}\nonumber\\
		&+\sum_{T_k\in\Phi_{ t}/b_{s}}\rho_{ t}HD_{ T_k}^{-\alpha_{ t}},\nonumber
	\end{align} 
	in which $D_{ N_i}$, $D_{ L_j}$ and $D_{ T_k}$ are the distances between the typical user and the interfering NLoS, LoS UAVs, and TBSs, respectively.
\end{definition}

	\subsection{Infrastructure Sharing}\label{sec_inf_sharing}
	We consider a scenario where EV operators share their infrastructure (EV charging stations) with UAV operators as far as their own services are qualified, and UAV operators pay for the corresponding service. Besides, if the infrastructure shared by EVs cannot maintain an acceptable network performance, say coverage probability is too low, UAVs need to install more dedicated charging stations, of which the density is denoted by $\Delta\lambda_{ c,d}$.
	
	The proposed objective functions are the total profit on the sides of UAVs' and EVs' operators, respectively. Note that in these two objective functions, the only constraints are the values of $\beta_{\{d,o\}}$, from 0 to the optimal values, where 0 denotes no sharing and optimal values are obtained from the maximal coverage probability (More details are provided in Section \ref{sec_num}).
\begin{definition}[Objective Function]\label{def_inf}
	For EVs' operator, the formulated function deals with the total extra waiting time of EVs and the profit paid by UAVs' operator per charging station per year: 
	\begin{align}
		C_{ e} &= w_{\rm wait}\Delta T_{\rm w, e v}\cdot 365\cdot 24\mu_e+w_{\rm i n f, e v} C_{\rm i n f, d}(\beta_{\{d,o\}}),\label{eq_Ce}
	\end{align}
where $w_{\rm i n f, ev}$, $w_{\rm wait}$ are  objective function coefficients (weights of objectives), $\Delta T_{\rm w, e v}\times 365\times 24\mu_e$ is the total extra waiting time of EVs per charging station and $C_{\rm i n f, d}(\beta_{\{d,o\}})$ is  the infrastructure sharing fee that UAVs' operators payed for EVs' operators. For example, if $w_{\rm wait}$ is larger than  $w_{\rm i n f, ev}$, it means that EVs' operator care more about the quality of their own services. 

For UAVs' operator, the formulated function deals with the improvement of network performance (coverage probability), cost of installing new charging stations and infrastructure sharing fee,
\begin{align}
		C_{u}&=w_{\rm c o v}\frac{P_{\rm c o v}^{'}(\Delta\lambda_{ c,d})}{P_{\rm c o v,in}}+\frac{w_{ c}\Delta\lambda_{ c,d}}{ \lambda_{ c,d}}+w_{\rm i n f, d}C_{\rm i n f, d}(\beta_{\{d,o\}}),\label{eq_Cu}
	\end{align}
where $w_{ c}$, $w_{\rm cov}$ and $w_{\rm inf, d}$  are  objective function coefficients, $P_{\rm c o v,in}$ and $P_{\rm c o v}^{'}(\Delta\lambda_{ c,d})$ are the initial performance (without infrastructure sharing), and performance with infrastructure sharing and more dedicated charging stations.
\end{definition}

	In the following text, we analyze the system performance from the perspective of UAVs and EVs. Since UAVs are for communication, we study the coverage probability, which is defined as the probability of the reference user being successfully served, e.g., SINR is greater than the predefined threshold. To do so, in Section \ref{sec_waitingtime} we compute the waiting time of UAVs under the aforementioned two association policies and charging station serving policies. We then in Section \ref{sec_availability} obtain the availability probability of UAVs, which is needed in coverage probability analysis, say (\ref{eq_Pcov}), and the final results of coverage probability is given in Section \ref{sec_pcov}. For EVs, we are concerned about the waiting time for charging. Hence, we in Section \ref{sec_waitingtime} compute the waiting time of EVs in the case of sharing charging stations.
	
\section{Waiting Time}\label{sec_waitingtime}
This section analyzes the waiting time of both EVs and UAVs and is the most important technical section of this paper. Notice that deriving the exact waiting time equations is tricky, our results are based on some reasonable assumptions and tight approximations. 
	
	Given that $T_{\rm ser}$ is much longer than $T_{\rm tra}$, therefore, we ignore $T_{\rm tra}$ when we analysis the waiting time and assume that all the drones have the same service time. Let $N_{d,ev}$ and $N_{d,d}$ be the number of drones recharging in EV charging stations and UAV charging stations, respectively.
	
	We first derive the waiting time of drones in their dedicated  charging stations.
	\begin{lemma}[Waiting Time of UAVs in UAV Charging Stations]
In the typical association cell of UAV charging station, conditioned on $N_{d,d}$, the waiting time $T_{w,d,d}$ is given by
\begin{align}
T_{\rm w,d,d|N_{d,d}} = T_{\rm ch,d,d}\bigg(N_{d,d}-\frac{T_{\rm ser}}{T_{\rm ch,d,d}}-1\bigg),
\end{align}		
where $T_{\rm ch,d,d}$ is the charging time of UAVs in their dedicated charging stations.
	\end{lemma}
	We then derive the waiting time of drones in EV charging stations based on the aforementioned serving priorities (Definition \ref{def_fifs} and Definition \ref{def_evfirst}). 
		
	\begin{figure}
		\centering
		\includegraphics[width = \columnwidth]{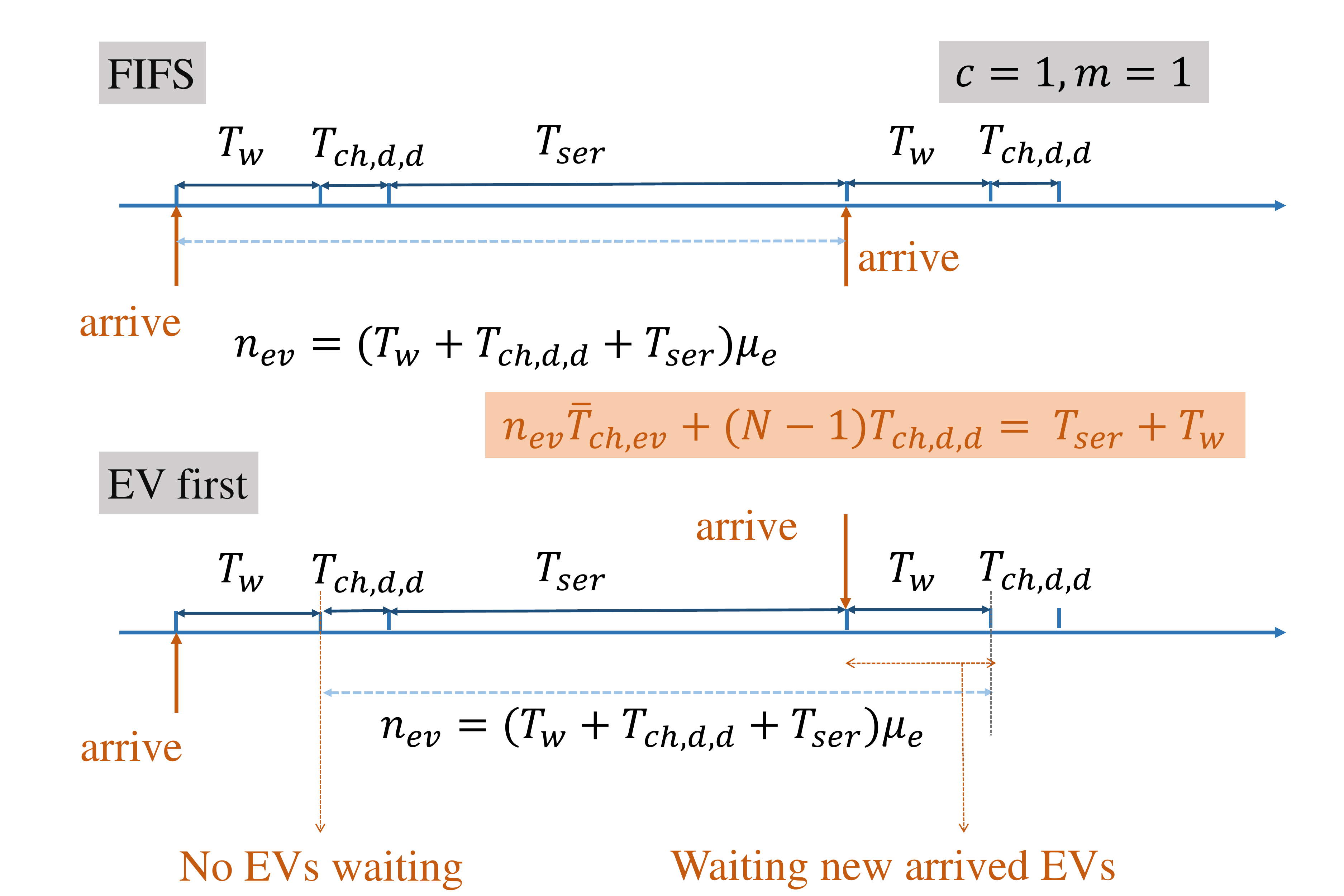}
		\caption{Illustration of the proof of waiting time (FIFS and EV first).}
		\label{fig_proof}
	\end{figure}
\begin{lemma}[Waiting Time of UAVs in EV Charging Stations (FIFS)]
	\label{lemma_FIFS}
	When the charging policy in the typical cell of EV charging station  is FIFS,  conditioned on the number of associated drones $N_{d,ev}$, the waiting time of UAVs is, in the case of $N_{d,ev}\geq mc(1+\frac{T_{\rm ser}}{T_{\rm ch,d,ev}})$, 
\begin{align}
	&T_{\rm w,d,ev|N_{d,ev}}= \nonumber\\
	&\frac{T_{\rm ch,d,ev}(\frac{N_{d,ev}}{mc}-1)+\frac{T_{\rm ch,d,ev}+T_{\rm ser}}{c}\mu_{\rm e}\mathbb{E}[T_{\rm ch,ev}]-T_{\rm ser}}{1-\frac{\mu_{e}\mathbb{E}[T_{\rm ch,ev}]}{c}}.
\end{align}
where $c$ is the capacity of EV charging stations (the number of EVs that can be served together) and $m$ denotes that $m$ UAVs can charge together within one charging slot. 
\end{lemma}
\begin{IEEEproof}
Here we provide the proof for $c = 1$, as shown in Fig. \ref{fig_proof}. From the perspective of a typical UAV, the waiting time can be computed as follows
\begin{align}
    (&T_{\rm w,d,ev\mid N_{d,ev}}+T_{\rm ch,d,ev}+T_{\rm ser})\mu_e\mathbb{E}[T_{\rm ch,ev}]\nonumber\\
    &+(N_{d,ev}-1)T_{\rm ch,d,ev}-T_{\rm ser} = T_{\rm w,d,ev\mid N_{d,ev}},
\end{align}
where the first term $(T_{\rm w,d,ev\mid N_{d,ev}}+T_{\rm ch,d,ev}+T_{\rm ser})\mu_e\mathbb{E}[T_{\rm ch,ev}]$ is the total charging time of EVs coming after the arrival of the typical UAV, $(N_{d,ev}-1)T_{\rm ch,d,ev}$ is the total charging time of other UAVs. The proof completes by simplifying the above equation.
\end{IEEEproof}
\begin{lemma}[Waiting time of UAVs in EV Charging Stations (EV First)]
	When EVs have higher priority while charging in EV charging stations, conditioned on the number of associated drones $N_{d,ev}$, the waiting time of UAVs is, in the case of $N_{d,ev}\geq mc(1+\frac{T_{\rm ser}}{T_{\rm ch,d,ev}})$, 
\begin{align}
	&T_{\rm w,d,ev|N_{d,ev}}= \nonumber\\
	 &\frac{T_{\rm ch,d,ev}(\frac{N_{d,ev}}{mc}-1)+\frac{T_{\rm ch,d,ev}+T_{\rm ser}}{c}\mu_{\rm e}\mathbb{E}[T_{\rm ch,ev}]-T_{\rm ser}}{1-\frac{\mu_{e}\mathbb{E}[T_{\rm ch,ev}]}{c}}.
\end{align}
\end{lemma}
\begin{IEEEproof}
Similar to the Proof of Lemma \ref{lemma_FIFS}, as shown in Fig. \ref{fig_proof}. From the perspective of a typical UAV, the waiting time is
\begin{align}
    (&T_{\rm w,d,ev\mid N_{d,ev}}+T_{\rm ch,d,ev}+T_{\rm ser})\mu_e\mathbb{E}[T_{\rm ch,ev}]\nonumber\\
&+(N_{d,ev}-1)T_{\rm ch,d,ev}-T_{\rm ser} = T_{\rm w,d,ev\mid N_{d,ev}}.
\end{align}
Here we find that the equations are exactly the same which is because of the scheduling of UAVs: UAVs visit the charging stations after a fixed amount of time frequently. If the UAV let the EVs charge first, there are no accumulated EVs waiting when the UAV start charging, however, it needs to wait the later arrival EVs charging during waiting in the charging stations.
\end{IEEEproof}

\begin{remark}\label{re_Tw}
	Note that since the arrival processes of UAVs are correlated in time and scheduled, the equations of waiting time of FIFS and EV first are exactly the same when $N_{d,ev}\geq mc(1+\frac{T_{\rm ser}}{T_{\rm ch,d,ev}})$. Besides, the charging time of UAVs in EV charging station is very long and	the density of UAVs is much higher than the density of charging stations, hence, $N_{d,ev}\geq cm(1+\frac{T_{\rm ser}}{T_{\rm ch,d,ev}})$ can be be satisfied  in the given system. While the waiting time of $N_{d,ev}< mc(1+\frac{T_{\rm ser}}{T_{\rm ch,d,ev}})$ is given in Appendix \ref{app_Tw_less}, it has a low probability. Therefore, in these two policies the waiting time of UAVs are approximately the same.  
\end{remark}

\begin{lemma}[Waiting Time of EVs]\label{lemma_TwEV}
	The waiting time of EVs are given by,
\begin{align}
	T_{\rm w,ev,fifs|N_{d,ev}} &= T_{\rm w,d,ev|N_{d,ev}}+T_{\rm w,ev,noDrone} + T_{\rm d|N_{d,ev}}, \nonumber\\
		T_{\rm w,ev,evfirst|N_{d,ev}} &= T_{\rm w,ev,noDrone} + T_{\rm d|N_{d,ev}},
\end{align}
where
\begin{align}
	T_{\rm w,ev,noDrone} &= \frac{\mu_{ e}\mathbb{E}^2[T_{\rm ch,ev}]}{1-\mu_{ e}\mathbb{E}^2[T_{\rm ch,ev}]}\frac{\mathbb{E}^2[T_{\rm ch,ev}]}{2\mathbb{E}[T_{\rm ch,ev}^2]},\nonumber\\
	T_{\rm d|N_{d,ev}} &= \min\bigg(1,\frac{N_{d,ev}T_{\rm ch,d,ev}}{mc(T_{\rm ch,d,ev}+T_{\rm ser})}\bigg)\nonumber\\
	&\int_{0}^{T_{\rm ch,d,ev}}\frac{c}{T_{\rm ch,d,ev}}(1-\frac{x}{T_{\rm ch,d,ev}})^{c-1}{\rm d}x.
\end{align}
\end{lemma}
\begin{IEEEproof}
	In the case of EV first, EVs don't need to wait for UAVs charging unless they are already charging. Hence, the waiting time of an EV is the sum of charging time of EVs come before it $T_{\rm w,ev,noDrone}$ and the remaining charging time of UAVs $T_{\rm d|N_{d,ev}}$.
	
	In the case of FIFS, assume that an EV arrives after the typical UAV, hence the waiting time of the EV is the sum of the waiting time of the UAV and the charging time of EVs before it and the remaining charging time of UAVs.
\end{IEEEproof}

\begin{remark}\label{re_EV_first}
	Based on Remark \ref{re_Tw} and Lemma \ref{lemma_TwEV}, we note that the waiting time of UAVs are approximately the same in these two serving policies while waiting time of EVs in the case FIFS is much longer than that of EV first. Besides, the waiting time of EVs increases slightly with the increase of $N_{ev}$ and then stays  constant, in the case of EV first. 
The same results shown in   Section \ref{sec_num}. Therefore, in the following analysis, we focus on analyzing the performance of infrastructure sharing based on the serving policy: EV first.
\end{remark}

\section{Availability Probability}\label{sec_availability}
This section derives the availability probability of UAVs based on two UAVs' association policies: (i) biased distance and (ii) independent thinning. Based on the association policies, we use the optimal $\beta_d$ and $\beta_o$ that maximize the coverage probability.

\subsection{Biased Distance}
In this section, we derive the $P_a$ under the biased distance policy: the typical UAV associates with the charging station based on $\min(R_{ s,ev},\beta_d R_{ s,d})$. To do so, we first formulate the distance distribution.
\begin{lemma}[Distance Distribution]
	The cumulative distribution function of the distances between the typical UAV and the nearest EV/UAV charging station denoted by $F_{ c,ev}(r)$ and $F_{ c,d}(r)$, respectively, are given by
	\begin{small}
\begin{align}
	F_{c,ev} &= 1-\exp\bigg(-2\pi\lambda_{ l}\int_{0}^{r}1-\exp(-2\lambda_{ p,ev}\sqrt{r^2-\rho^2}){\rm d}\rho\bigg),\\
	F_{c,d} &= 1-\exp\bigg(-2\pi\lambda_{ l}\int_{0}^{r}1-\exp(-2\lambda_{ p,d}\sqrt{r^2-\rho^2}){\rm d}\rho\bigg),
\end{align}
\end{small}
taking the derivative, their PDF are, respectively, given by
\begin{align}
&f_{ c,ev}(r) = 2\pi\lambda_{ l}\int_{0}^{r}\frac{2\lambda_{ p,ev}r}{\sqrt{r^2-\rho^2}}\exp(-2\lambda_{ p,ev}\sqrt{r^2-\rho^2}){\rm d}\rho\nonumber\\
&\exp\bigg(-2\pi\lambda_{ l}\int_{0}^{r}1-\exp(-2\lambda_{ p,ev}\sqrt{r^2-\rho^2}){\rm d}\rho\bigg),\\
&f_{ c,d}(r) = 2\pi\lambda_{ l}\int_{0}^{r}\frac{2\lambda_{ p,d}r}{\sqrt{r^2-\rho^2}}\exp(-2\lambda_{ p,d}\sqrt{r^2-\rho^2}){\rm d}\rho\nonumber\\
&\exp\bigg(-2\pi\lambda_{ l}\int_{0}^{r}1-\exp(-2\lambda_{ p,d}\sqrt{r^2-\rho^2}){\rm d}\rho\bigg).
\end{align}
\end{lemma}

The probability that the cluster UAV associates with the typical charging station is a function of distance and association weights. In the following lemma, we identify the conditional and unconditional association probability.
\begin{lemma}[UAV's Association Probability]\label{lemm_uav_ass}
Given that the nearest serving charging station is at $r$ away, the association probability is given by,
\begin{align}
	\mathcal{A}_{ev\mid r}(r,\beta_d) &= \mathbb{P}(R_{ s,ev}<\beta_dR_{ s,d})\nonumber\\
	&=\mathbb{P}(\frac{r}{\beta_d}<R_{ s,d})= \bar{F}_{c,d}(\frac{r}{\beta_d}),\\
	\mathcal{A}_{d\mid r}(r,\beta_d) &= \mathbb{P}(R_{ s,ev}>\beta_dR_{ s,d})\nonumber\\
	&=\mathbb{P}(R_{ s,ev}>\beta_d r)= \bar{F}_{c,ev}(\beta_d r),
\end{align}
taking the expectation over the distance, the probabilities of associating with EV/UAV charging stations are
\begin{align}
    \mathcal{A}_{ev}(\beta_d) &= \int_{0}^{\infty}\bar{F}_{c,d}(\frac{r}{\beta_d})f_{c,ev}(r){\rm d}r,\nonumber\\
	\mathcal{A}_{d}(\beta_d) &= \int_{0}^{\infty}\bar{F}_{c,ev}(r\beta_d)f_{c,d}(r){\rm d}r.
\end{align}
\end{lemma}
Note that for a PPP, the probability of the number of points filling in a certain cell is proportional to the area of this cell. Hence in the following lemma, we provide the area approximation. Here we adopt both the fitting and approximation formulas of cell area based on Poisson Voronoi (PV) tessellations, however, it shows that they are both tight for CV and MWCV tessellations, in the case that line density is large while point density is low.
\begin{lemma}[Area Approximation]
Using the approximation mentioned in \cite{singh2013offloading}, the association area of EV/UAV charging stations can be approximated as
\begin{align}
f_{C}(c) &= \frac{b^a}{\Gamma(a)}(\frac{\lambda}{\mathcal{A}(\beta_d)})(\frac{\lambda}{\mathcal{A}(\beta_d)}c)^{a-1}\exp(-b\frac{\lambda}{\mathcal{A}(\beta_d)}c),
\end{align}
where $\lambda$ and $\mathcal{A}(\beta_d)$ are the density of charging stations ($\lambda_{c,ev}$ or $\lambda_{c,d}$) and UAV's association probability defined in Lemma \ref{lemm_uav_ass} ($\mathcal{A}_{c,ev}(\beta_d)$ or $\mathcal{A}_{c,d}(\beta_d)$). As a random UAV is more likely to lie in a larger cell, which is know as biased cell, and the PDF of the biased cell is
\begin{align}
	f_{C^{'}}(c) &= \frac{b^a}{\Gamma(a)}(\frac{\lambda}{\mathcal{A}(\beta_d)})(\frac{\lambda}{\mathcal{A}(\beta_d)}c)^{a}\exp(-b\frac{\lambda}{\mathcal{A}(\beta_d)}c).
\end{align}
\end{lemma}

Knowing the area of the biased cell, apart from the typical UAV, the PMF of the number of other UAVs are given in the following lemma.
\begin{lemma}[Average Number of UAVs]
The PMF of the other UAVs associated with the typical EV/UAV charging stations are given by
\begin{align}
	\mathbb{P}(N_{d,ev}=n) =& \frac{\Gamma(a+n+1)}{\Gamma(a)} \frac{b^{a}}{n !} (\frac{\lambda_{ c,ev}}{\mathcal{A}_{ev}(\beta_d)})^{a+1} \nonumber\\
	&\times\frac{\lambda_{\mathrm{u}}^{n}}{\left(b \frac{\lambda_{ c,ev}}{\mathcal{A}_{ev}(\beta_d)}+\lambda_{\mathrm{u}}\right)^{a+n+1}},\nonumber\\
	\mathbb{P}(N_{d,d}=n) =& \frac{\Gamma(a+n+1)}{\Gamma(a)} \frac{b^{a}}{n !} (\frac{\lambda_{ c,d}}{\mathcal{A}_{d}(\beta_d)})^{a+1} \nonumber\\
	&\times \frac{\lambda_{\mathrm{u}}^{n}}{\left(b \frac{\lambda_{ c,d}}{\mathcal{A}_{d}(\beta_d)}+\lambda_{\mathrm{u}}\right)^{a+n+1}}.
\end{align}
\end{lemma}
\begin{IEEEproof}
	The number of UAVs per CV cell is a Poisson random variable, with parameter cell area, given by
	\begin{small}
	\begin{align}
	&\mathbb{P}(N=n) =\mathbb{E}_{C^{\prime}}\left[\mathbb{P}\left(N=n \mid C^{\prime}\right)\right] =\int_{0}^{\infty} \mathbb{P}(N=n) f_{C^{\prime}}(c) \mathrm{d} c \nonumber\\
	&=\int_{0}^{\infty} \frac{\left(\lambda_{\mathrm{u}} c\right)^{n} e^{-\lambda_{\mathrm{u}} c}}{n !} \frac{b^a}{\Gamma(a)}(\frac{\lambda}{\mathcal{A}(\beta_d)})(\frac{\lambda}{\mathcal{A}(\beta_d)}c)^{a}\exp(-b\frac{\lambda}{\mathcal{A}(\beta_d)}c) {\rm d} c \nonumber\\
	&=\frac{\Gamma(a+n+1)}{\Gamma(a)} \frac{b^{a}}{n !} (\frac{\lambda}{\mathcal{A}(\beta_d)})^{a+1} \frac{\lambda_{\mathrm{u}}^{n}}{\left(b \frac{\lambda}{\mathcal{A}(\beta_d)}+\lambda_{\mathrm{u}}\right)^{a+n+1}}.\label{eq_N}
\end{align}
\end{small}
\end{IEEEproof}
Notice that conditioned on associating with the tagged charging station, the distance distribution is different from the first contact distance in PLCP, since it is also influenced by nearby charging stations and association weight.
\begin{lemma}[Conditional Distance Distribution]
	Let $Y_{\{ev,d\}}$ be the distance between the typical UAV and its serving EV/UAV charging station. Conditioned on association, the PDF of $Y_{\{ev,d\}}$ is 
	\begin{align}
f_{Y_{ev}}(y) &=  \frac{\bar{F}_{c,d}(\frac{y}{\beta_d})f_{c,ev}(y)}{\mathcal{A}_{ev}(\beta_d)},\\
f_{Y_{d}}(y) &=  \frac{\bar{F}_{c,ev}(y\beta_d)f_{c,d}(y)}{\mathcal{A}_{d}(\beta_d)}.
	\end{align}
\end{lemma}
\begin{IEEEproof}
	$Y_{ev}$ has the same distance distribution as $R_{s,ev}$ conditioned on the typical UAV being associated with the EV charging station,
\begin{align}
&\mathbb{P}(Y_{ev}>y) \nonumber\\
&= \frac{\mathbb{P}(R_{ s,ev}>y \mid \text{associate with EV charging station})}{\mathbb{P}(\text{associate with EV charging station})} \nonumber\\
&= \frac{\int_{y}^{\infty} \bar{F}_{c,d}(\frac{r}{\beta_d})f_{ev}(r){\rm d}r}{\mathcal{A}_{ev}(\beta_d)}, 
\end{align}
proof completes by taking the derivative.
\end{IEEEproof}
Now we proceed to present the conditional and unconditional availability probability of UAVs in the case of biased distance association policy.
\begin{theorem}[Availability Probability]\label{the_Pa_bias}
	Availability probability under the presented UAVs' association policy is given by
\begin{align}
	&P_{\rm a,bias}= \sum_{n = 0}^{\infty}\bigg(\int_{0}^{\frac{vB_{\rm max}}{2p_m}}g_{ev}(y\mid n)\bar{F}_{c,d}(\frac{y}{\beta_d})f_{ev}(y){\rm  d}y \nonumber\\
	 &\quad\times\frac{\Gamma(a+n+1)}{\Gamma(a)} \frac{b^{a}}{n !} (\frac{\lambda_{ c,ev}}{\mathcal{A}_{ev}(\beta_d)})^{a+1} \frac{\lambda_{\mathrm{u}}^{n}}{\left(b \frac{\lambda_{ c,ev}}{\mathcal{A}_{ev}(\beta_d)}+\lambda_{\mathrm{u}}\right)^{a+n+1}}\bigg)\nonumber\\
	&+ \sum_{n = 0}^{\infty}\bigg(\int_{0}^{\frac{vB_{\rm max}}{2p_m}}g_{d}(y\mid n)\bar{F}_{c,ev}(y\beta_d)f_{d}(y){\rm  d}y \nonumber\\
	&\quad\times\frac{\Gamma(a+n+1)}{\Gamma(a)} \frac{b^{a}}{n !} (\frac{\lambda_{ c,d}}{\mathcal{A}_{d}(\beta_d)})^{a+1} \frac{\lambda_{\mathrm{u}}^{n}}{\left(b \frac{\lambda_{ c,d}}{\mathcal{A}_{d}(\beta_d)}+\lambda_{\mathrm{u}}\right)^{a+n+1}}\bigg),
\end{align}
where $g_{ev}(y\mid n) = \frac{vB_{\rm max}-2yp_{m}}{vB_{\rm max}-2y(p_{m}-p_{s})+vp_{s}(T_{\rm ch,d,ev}+T_{\rm w,d,ev\mid N_{d,ev}})}$ and $g_{d}(y\mid n) = \frac{vB_{\rm max}-2yp_{m}}{vB_{\rm max}-2y(p_{m}-p_{s})+vp_{s}(T_{\rm ch,d,d}+T_{ w,d,d\mid N_d})}$.	
\end{theorem}
\begin{IEEEproof}
	See Appendix \ref{app_Pa_bias}.
	\end{IEEEproof}
\subsection{Independent Thinning}
In this section, we derive the $P_a$ under the independent thinning policy: the typical UAV associates with the charging station based on the independent thinning.

In this policy, we consider UAVs are divided into two parts: associating with $\Phi_{ c,ev}$ and associating with $\Phi_{ c,d}$ with probability $\beta_o$ and $1-\beta_o$, respectively. Following the same steps in (\ref{eq_N}) and substituting the UAVs' density $\beta_o\lambda_{ u}$ and $(1-\beta_o)\lambda_{ u}$ and association probabilities here equal to 1, we derive the availability probability $P_{a,th}$.
\begin{theorem}[Availability Probability]\label{the_Pa_thi}
Availability probability under the independent thinning association policy is 
	\begin{align}
		P_{\rm a,th} &=\beta_o\sum_{n = 0}^{\frac{vB_{\rm max}}{2p_m}}\bigg(\int_{0}^{\infty}g_{ev}(y\mid n)f_{c,ev}(y){\rm  d}y\nonumber\\
		&\quad\times\frac{\Gamma(a+n+1)}{\Gamma(a)} \frac{b^{a}}{n !} \frac{\bigg(\frac{\beta_o\lambda_{\mathrm{u}}}{\lambda_{ c,ev}}\bigg)^{n}}{\left(b +\frac{\beta_o\lambda_{\mathrm{u}}}{\lambda_{ c,ev}}\right)^{a+n+1}}\bigg)\nonumber\\
		&+ (1-\beta_o)\sum_{n = 0}^{\frac{vB_{\rm max}}{2p_m}}\bigg(\int_{0}^{\infty}g_{d}(y\mid n)f_{c,d}(y){\rm  d}y\nonumber\\
		&\quad\times\frac{\Gamma(a+n+1)}{\Gamma(a)} \frac{b^{a}}{n !}  \frac{\bigg(\frac{(1-\beta_o)\lambda_{\mathrm{u}}}{\lambda_{ c,d}}\bigg)^{n}}{\left(b +\frac{(1-\beta_o)\lambda_{\mathrm{u}}}{\lambda_{ c,d}}\right)^{a+n+1}}\bigg).
	\end{align}
\end{theorem}
\begin{IEEEproof}
    Similar to Proof of Theorem \ref{the_Pa_bias}, thus omitted here.
\end{IEEEproof}
Note that in these two association policies, the expectation of distance under the biased distance policy is shorter than the other one. Hence, its performance is expected to be slightly better than the other one. More details are shown in Section \ref{sec_num}.

In the following part of the paper, we assume that the optimal value of $\beta_o$ and $\beta_d$ are used to maximize the coverage probability given in the next part, and to simplify the notation we use $P_a$ as the optimal value for both $P_{\rm a,bias}$ and $P_{\rm a,th}$.

	\section{Coverage Probability}
\label{sec_pcov}
	
	In this section, our goal is to analyze the coverage probability. To do so, the distance between the BSs and users are required and given in the following lemma. Recall that $\Phi_{ u^{'}}$ is the set formed by available UAVs from the original point process $\Phi_{ u}$, with thinning probability $P_a$, the density of available UAVs is $\lambda_{ u}^{'}=P_{ a}\lambda_{ u}$.
	\begin{lemma}[Distance Distribution]
		The probability density function of the distances between the typical user and the cluster UAV, the nearest available NL/LoS UAV, and the nearest TBS, denoted by $f_{ R_{u_o}}(r)$, $f_{ R_{u^{'}_n}}(r)$, $f_{ R_{u^{'}_l}}(r)$ and $f_{ R_t}(r)$, are  respectively given by
	\begin{align}
		 &f_{ R_{u_o}}(r) = \frac{2r}{r_{c}^2}, \quad h\leq r \leq \sqrt{r_c^2+h^2},\label{dist_u_o}\\
		&f_{ R_{u^{'}_n}}(r) = 2\pi\lambda_{ u}^{'}P_{ n}(r)r\nonumber\\
		&\quad\times\exp\bigg(-2\pi\lambda_{ u}^{'}\int_{0}^{\sqrt{r^2-h^2}}zP_{ n}(\sqrt{z^2+h^2}){ d}z\bigg),\label{dist_u_p_n}\\
		&f_{ R_{u^{'}_l}}(r) = 2\pi\lambda_{ u}^{'}P_{ l}(r)r\nonumber\\
		&\quad\times\exp\bigg(-2\pi\lambda_{ u}^{'}\int_{0}^{\sqrt{r^2-h^2}}zP_{ l}(\sqrt{z^2+h^2}){ d}z\bigg)\label{dist_u_p_l},\\
		&f_{ R_t}(r) = 2\pi r\lambda_{t}\exp(-\pi\lambda_{t}r^2),
	\end{align}
	where $P_{ n}(r)$ and $P_{ l}(r)$ are defined in (\ref{pl_pn}).
	\end{lemma}

Recall that we assume that the typical user associates with the cluster UAV if it is available, otherwise, it associates with a nearby available UAV or TBS based on the average received power. The following lemma gives the association probability of the typical user.

\begin{lemma}[Association Probability]\label{lem_user_ass}
Given the serving BS located at $r$ away, the association probabilities of the typical user are, respectively, given by
\begin{align}
\mathcal{A}_{\rm LoS}(r) &= \exp\bigg(-2\pi\lambda_{u}^{'}\int_{0}^{\sqrt{d_n^2(r)-h^2}}zP_n(\sqrt{z^2+h^2}){\rm d}z\bigg)\nonumber\\
&\quad\times\exp\bigg(-2\pi\lambda_{t}d_{lt}^2(r)\bigg),\\
\mathcal{A}_{\rm NLoS}(r) &=\exp\bigg(-2\pi\lambda_{u}^{'}\int_{0}^{\sqrt{d_l^2(r)-h^2}}zP_l(\sqrt{z^2+h^2}){\rm d}z\bigg)\nonumber\\
&\quad\times\exp\bigg(-2\pi\lambda_{t}d_{nt}^2(r)\bigg),\\
\mathcal{A}_{\rm TBS}(r) &= \exp\bigg(-2\pi\lambda_{u}^{'}\int_{0}^{\sqrt{d_{tl}^{2}(r)-h^2}}zP_l(\sqrt{z^2+h^2}){\rm d}z\bigg)\nonumber\\
&\times\exp\bigg(-2\pi\lambda_{u}^{'}\int_{0}^{\sqrt{d_{tn}^2(r)-h^2}}zP_n(\sqrt{z^2+h^2}){\rm d}z\bigg),
\end{align}
where $d_{lt}(r) = (\frac{\rho_t}{\rho_u\eta_l  })^{\frac{1}{\alpha_t}}r^{\frac{\alpha_l}{\alpha_t}}$, $d_{n}(r) = \max\bigg( (\frac{\eta_n}{\eta_l})^{\frac{1}{\alpha_n}} r^{\frac{\alpha_l}{\alpha_n}},h\bigg)$, $d_l(r) = \max\bigg(h,(\frac{\eta_l}{\eta_n})^{\frac{1}{\alpha_l}} r^{\frac{\alpha_n}{\alpha_l}}\bigg)$, $d_{nt}(r) = (\frac{\rho_t}{\rho_u\eta_n  })^{\frac{1}{\alpha_t}}r^{\frac{\alpha_n}{\alpha_t}}$, $d_{tl}(r) = \max\bigg(h,(\frac{\rho_u\eta_l}{\rho_t})^{\frac{1}{\alpha_l}} r^{\frac{\alpha_t}{\alpha_l}}\bigg)$ and $d_{tn}(r) = \max\bigg(h,(\frac{\rho_u\eta_n}{\rho_t})^{\frac{1}{\alpha_n}} r^{\frac{\alpha_t}{\alpha_n}}\bigg)$.
\end{lemma}
\begin{IEEEproof}
	See Appendix \ref{app_user_ass}.
\end{IEEEproof}

Laplace transform of the aggregate interference is the final requirement to the coverage probability.
\begin{lemma}[Laplace Transform]\label{lem_laplace}
	Given the serving BS $b_s$, the Laplace transform of the interference is given by
	\begin{align}
&\mathcal{L}_{I}(s,r) \nonumber\\
=&\exp\biggl(-2\pi\lambda_{ u}^{'}\int_{a(\|x\|)}^{\infty}\bigg[1-\bigg(\frac{m_{ n}}{m_{ n}+s\eta_{ n}\rho_{ u}(z^2+h^2)^{-\frac{\alpha_{ n}}{2}}}\bigg)^{m_{ n}}\bigg]\nonumber\\
&\quad\times zP_{ n}(\sqrt{z^2+h^2}){ d}z\biggl)\nonumber\\
&\times \exp\biggl(-2\pi\lambda_{ u}^{'}\int_{b(\|x\|)}^{\infty}\bigg[1-\bigg(\frac{m_{ l}}{m_{ l}+s\eta_{ l}\rho_{ u}(z^2+h^2)^{-\frac{\alpha_{ l}}{2}}}\bigg)^{m_{ l}}\bigg]\nonumber\\
&\quad\times zP_{ l}(\sqrt{z^2+h^2}){\rm d}z\biggl)\nonumber\\
&\times \exp\biggl(-2\pi\lambda_{ t}\int_{c(\|x\|)}^{\infty}\bigg[1-(\frac{1}{1+s\rho_{ t}z^{-\alpha_{ t}}})\bigg]z{\rm d}z\biggl),
	\end{align}
	in which,
\begin{align}
	&a(\|x\|)=\left\{ 
	\begin{aligned}
		0,  & \quad \text{\rm if} \quad bs \in \Phi_{ u_{o}},\\
		\sqrt{d_{n}^2(\|x\|)-h^2},  & \quad \text{\rm if} \quad bs \in \Phi_{ u^{'}_{l}},\\
		\sqrt{\|x\|^2-h^2},  & \quad \text{\rm if} \quad bs \in \Phi_{ u_{n}^{'}},\\
		\sqrt{d_{tn}^2(\|x\|)-h^2},  & \quad \text{\rm if} \quad bs \in \Phi_{ t},\\
	\end{aligned} \right.\nonumber\\
&b(\|x\|)=\left\{ 
	\begin{aligned}
		0,  & \quad \text{\rm if} \quad bs \in \Phi_{ u_{o}},\\
		\sqrt{\|x\|^2-h^2},  & \quad \text{\rm if} \quad bs \in \Phi_{ u^{'}_{l}},\\
		\sqrt{d_{l}^2(\|x\|)-h^2},  & \quad \text{\rm if} \quad bs \in \Phi_{ u_{n}^{'}},\\
		\sqrt{d_{tl}^2(\|x\|)-h^2},  & \quad \text{\rm if} \quad bs \in \Phi_{ t},\\
	\end{aligned} \right.\nonumber\\
&c(\|x\|)=\left\{ 
	\begin{aligned}
		0,  & \quad \text{\rm if} \quad bs \in \Phi_{ u_{o}},\\
		d_{lt}(\|x\|),  & \quad \text{\rm if} \quad bs \in \Phi_{ u^{'}_{l}},\\
		d_{nt}(\|x\|),  & \quad \text{\rm if} \quad bs \in \Phi_{ u_{n}^{'}},\\
		\|x\|,  & \quad \text{\rm if} \quad bs \in \Phi_{ t},\\
	\end{aligned} \right.\nonumber
\end{align}
\end{lemma}	
\begin{IEEEproof}
	The Laplace transform is computed using moment generating function (MGF) of Gamma distribution, probability generation functional (PGFL) of inhomogeneous PPP. For more details,  please refer to \cite{9444343}.
\end{IEEEproof}

Using the results derived thus far, the total coverage probability can be obtained as given in the following theorem.
\begin{theorem}[Coverage Probability]\label{theo_cov}
When the typical user is associated with the cluster UAV, conditioned on LoS or NLoS, the coverage probability is
\begin{align}
		\label{pro_eq_P_cov_Uo_L}
P_{\rm cov,U_{o,l}} &= \int_{h}^{\sqrt{h^2+r_{c}^{2}}}\sum_{k=0}^{m_{ l}-1}\frac{(-m_{ l}g_{ l}(r))^k}{k!}\nonumber\\
&\times\frac{\partial^{k}}{\partial s^{k}}\bigg[\mathcal{L}_{\sigma^2+I}(s,r)\bigg]_{s=m_{ l}g_{ l}(r)}P_{ l}(r) \frac{2r}{r_{c}^2}{\rm d}r,\\
P_{\rm cov,U_{o,n}}&= \int_{h}^{\sqrt{h^2+r_{c}^{2}}}\sum_{k=0}^{m_{ n}-1}\frac{(-m_{ n}g_{ n}(r))^k}{k!} \nonumber\\
&\times\frac{\partial^{k}}{\partial s^{k}}\bigg[\mathcal{L}_{\sigma^2+I}(s,r)\bigg]_{s=m_{ n}g_{ n}(r)}P_{ n}(r) \frac{2r}{r_{c}^2}{\rm d}r.
\end{align}
	where $	g_{ l}(r) = \gamma(\eta_{l}\rho_u)^{-1}r^{\alpha_{ l}}$ and $g_{ n}(r)=\gamma(\eta_{n}\rho_u)^{-1}r^{\alpha_{ n}}$.
	
When the cluster UAV is unavailable, the coverage probability when associating with the nearest LoS/NLoS available UAV $P_{\rm cov,\hat{U}_o,l}$ and $P_{\rm cov,\hat{U}_o,n}$ can be given by
\begin{align}
	\label{pro_eq_P_cov_Up_L}
	P_{\rm cov,\hat{U}_o,l}&=\int_{h}^{\infty}\mathcal{A}_{\rm LoS}(r)\sum_{k=0}^{m_{ l}-1}\frac{(-m_{ l}g_{ l}(r))^k}{k!}\nonumber\\
	&\times\frac{\partial^{k}}{\partial s^{k}}\mathcal{L}_{\sigma^2+I}(s,r)|_{s=m_{ l}g_{ l}(r)}\bigg]f_{ R_{u^{'},l}}(r) {\rm d}r,\\
	P_{\rm cov,\hat{U}_o,n} &= \int_{h}^{\infty}\mathcal{A}_{\rm NLoS}(r)\sum_{k=0}^{m_{ n}-1}\frac{(-m_{ n}g_{ n}(r))^k}{k!}\nonumber\\
	&\times\frac{\partial^{k}}{\partial s^{k}}\mathcal{L}_{\sigma^2+I}(s,r)|_{s=m_{ n}g_{ n}(r)}\bigg]f_{ R_{u^{'},n}}(r) {\rm d}r,	\label{pro_eq_P_cov_Up_n}
\end{align}
where $f_{ R_{u^{'},l}}(r)$ is given in (\ref{dist_u_p_l}). 

When the typical UAV is unavailable, the coverage probability of associating with the nearest TBS $P_{\rm cov,t}$ can be written as
\begin{align}
	P_{\rm cov,t} &= \int_{0}^{+\infty}\mathcal{L}_{\sigma^2+I}(s,r)|_{s = \theta\rho_t^{-1} r^{\alpha_{ t}}}\mathcal{A}_{\rm TBS}(r)f_{ R_t}(r){\rm d}r.
\end{align}

\end{theorem}
\begin{IEEEproof}
The coverage probability is derived by the fact that (i) the uniform distribution of the users in the disk with radius $r_c$ and $g_{ l}(r)=\frac{\gamma \sigma^2}{\eta_{ l} r^{-\alpha_{ l}}\rho_{ u}}$, (ii) the definition: $\Bar{F}_{ G}(g)=\frac{\Gamma_{u}(m,g)}{\Gamma(m)}$, where $\Gamma_{u}(m,g)=\int^{\infty}_{mg}t^{m-1}e^{-t}dt$ is the upper incomplete Gamma function, and (iii) the definition $\frac{\Gamma_{u}(m,g)}{\Gamma(m)}=\exp(-g)\sum^{m-1}_{k=0}\frac{g^{k}}{k!}$.
\end{IEEEproof}
	\begin{table*}\caption{Table of Parameters}\label{table_par}
	\centering
	\begin{center}
		\resizebox{1.6\columnwidth}{!}{
			\renewcommand{\arraystretch}{1}
			\begin{tabular}{ {c} | {c} | {c}  }
				\hline
				\hline
				\textbf{Parameter} & \textbf{Symbol} & \textbf{Simulation Value}  \\ \hline
				Line density, point density & $\lambda_{ l}$; $\lambda_{ p,ev}$, $\lambda_{ p,d}$ & $24/\pi\times 10^{-3}$; $2.1\times 10^{-5}$, $1.05\times 10^{-5}$ \\ \hline
				Charging station density & $\lambda_{ c,d}$, $\lambda_{ c,ev}$ & $0.5$/km$^{2}$, $0.25$/km$^{2}$ \\ \hline
				TBS density, UAV density & $\lambda_{ t}$, $\lambda_{ u}$ & $1$/km$^{2}$, $4$/km$^{2}$ \\ \hline
				Average and deviation of SOC & $\mu$, $\sigma$ & 3 to 0.6\\ \hline
				EV charging rate & $P_{\rm cha}$ & 120 kw \\ \hline
				CV (or MWCV) tessellation fitting parameters & $a,b$ & 3.5\\ \hline
				Traveling-related power & $p_{ m}$ & 161.8 W \\\hline
				Service-related power & $p_{ s}$ & 177.5 W \\\hline
				UAV altitude & $h$ & 100 m\\\hline
				UAV velocity & $v$ & 18.46 m/s \\\hline
				Battery capacity & $B_{\rm max}$ & 177.6 W$\cdot$H \\\hline
				Charging time & $T_{\rm ch,d,ev}$, $T_{\rm ch,d,d}$ & 30, 5 min\\\hline
				Radius of MCP disk & $r_c$ & 100 m \\\hline
				N/LoS environment variable & $c_1, c_2$ & 25.27,0.2 \\\hline
				Transmission power & $\rho_{ u}$, $\rho_{ t}$ & 0.2 W, 10 W \\\hline
				SINR threshold & $\gamma$ & 0 dB \\\hline
				Noise power & $\sigma^2_n $ & $10^{-9}$ W\\\hline
				N/LoS and active charging station path-loss exponent & $\alpha_{ n},\alpha_{ l},\alpha_{ t}$ & $4,2.1,4$ \\\hline
				N/LoS fading gain & $m_{ n},m_{ l}$ & $1,3$ \\\hline
				N/LoS additional loss& $\eta_{ n},\eta_{ l}$ & $-20,0$ dB \\\hline
				Cost & $c_{\rm vol},c_{\rm main}$ & $0.2$ USD/kwh, $1131$ USD/year \\\hline
				Weight & $w_{\rm wait},w_{\rm inf,ev}$, $w_{\rm cov},w_{ c}$ & (-1/3,2/3) or (-2/3,1/3),   (8,1) or (6,1) 
				\\\hline\hline
		\end{tabular}}
	\end{center}
\end{table*}
It can be seen that the above coverage probability equations require evaluating higher order of derivatives of the Laplace transform. Using the upper bound of the CDF of the Gamma distribution \cite{alzer1997some}, the above equations can be approximated in the following lemma.


\begin{lemma}[Approximated Coverage Probability]\label{lem_app_Pcov}
Following the steps in \cite{alzenad2019coverage} and \cite{bai2014coverage}, the approximate coverage probabilities are given by,
\begin{align}
&P_{\rm cov,U_{o,\{l,n\}}} = \sum_{k=1}^{m_{ \{l,n\}}}\binom{m_{ \{l,n\}}}{k}(-1)^{k+1}\int_{h}^{\sqrt{h^2+r_{c}^2}}\nonumber\\
&\quad\mathcal{L}_{ \sigma^2+I}(k\beta_{2}(m_{ \{l,n\}})m_{ \{l,n\}}g_{ \{l,n\}}(r),r)P_{ \{l,n\}}(r)\frac{2r}{r_c^2}{\rm d}r,\\
&P_{\rm cov,\hat{U}_{o,\{l,n\}}}= \sum_{k=1}^{m_{ \{l,n\}}}\binom{m_{ \{l,n\}}}{k}(-1)^{k+1}\int_{h}^{\infty}\mathcal{A}_{\rm \{L,NL\}oS}(r)\nonumber\\
&\times f_{ R_{u^{'}_{\{l,n\}}}}(r)\mathcal{L}_{ \sigma^2+I}(k\beta_{2}(m_{ \{l,n\}})m_{ \{l,n\}}g_{ \{l,n\}}(r),r){\rm d}r,
\end{align}
in which $\beta_2 = (m_{\{l,n\}}!)^(-\frac{1}{m_{\{l,n\}}})$.
\end{lemma}
\begin{IEEEproof}
The approximation  is derive by using the definition of CCDF of Gamma function, upper imcomplete Gamma function and Binomial theorem.
\end{IEEEproof}
 
	\section{Infrastructure Sharing}
 
\begin{figure*}[ht]
	\centering
	\includegraphics[width = 2\columnwidth]{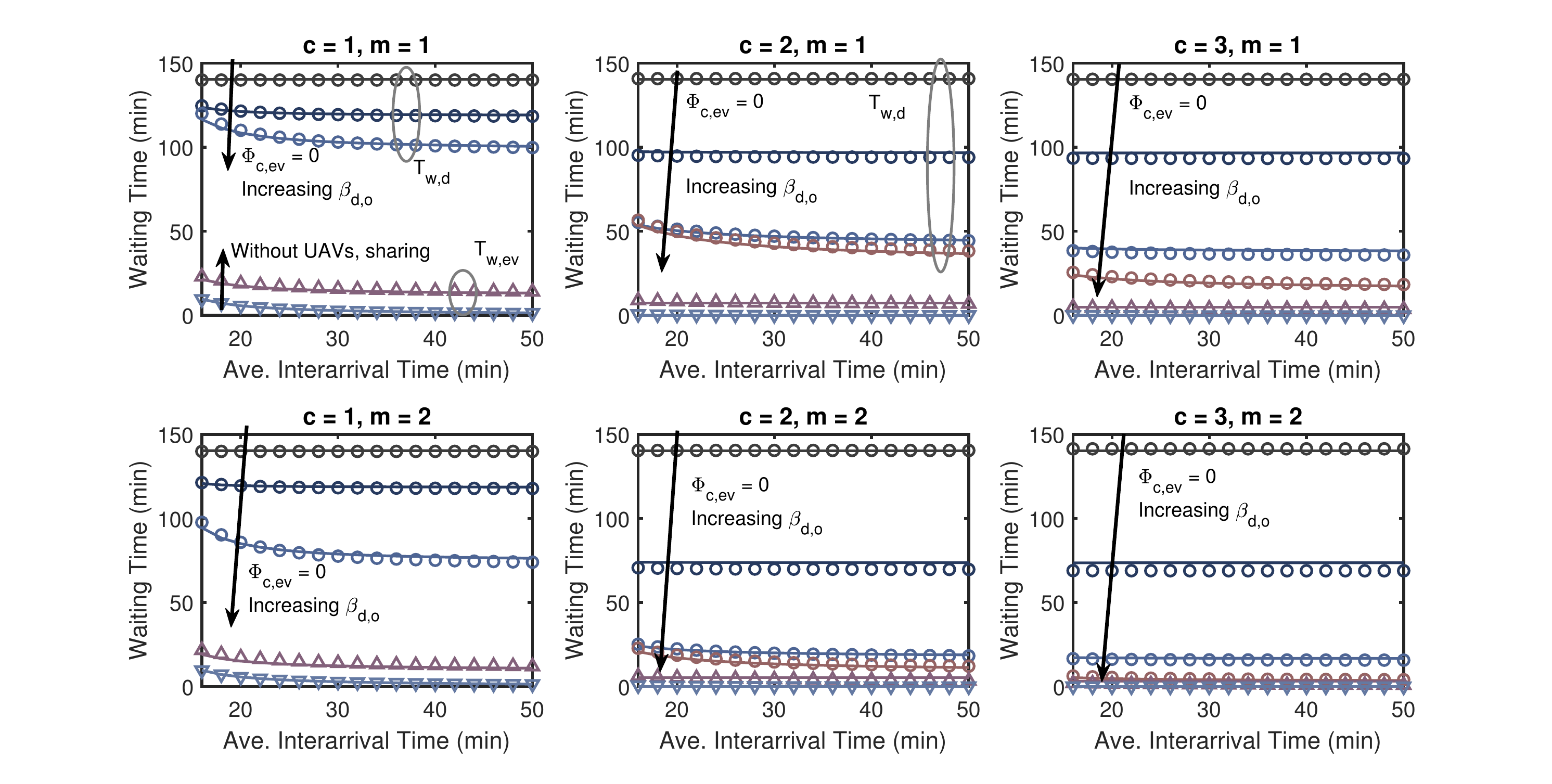}
	\caption{Analysis and simulation results of waiting time of UAVs under different average interarrival time of EVs in the case of sharing and no sharing. $c$ and $m$ are charging slots and the number of UAVs that can be charged together in EV charging stations. The dash lines are the waiting time of UAVs in their own charging stations without infrastructure sharing, and the sold lines are with the infrastructure sharing: increasing $\beta_{\{d,o\}}$ (the direction of the arrows) until the optimal value (maximize the coverage probability).}
	\label{fig_Tw}
\end{figure*}
\begin{figure*}[ht]
	\centering
	\includegraphics[width = 2\columnwidth]{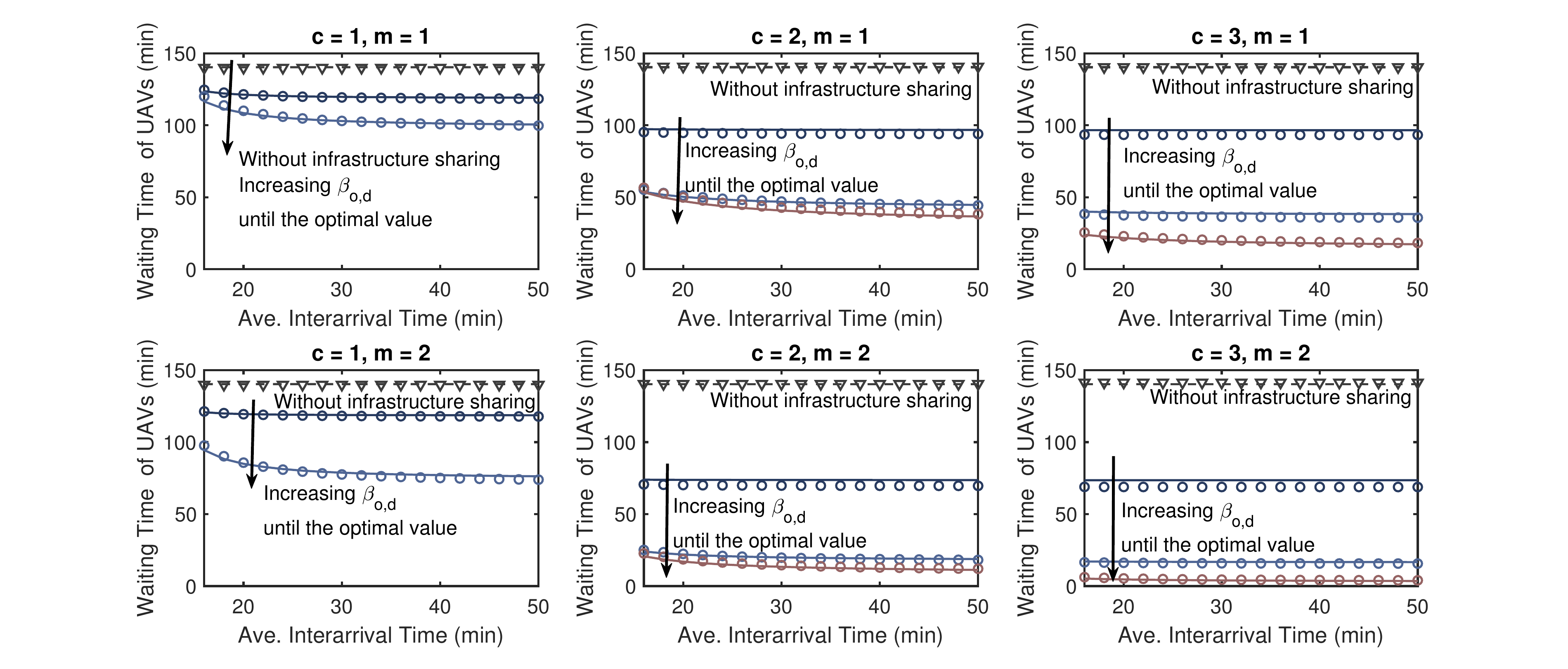}
	\caption{Analysis and simulation results of waiting time of EVs under different average interarrival time in the case of sharing and no sharing. $c$ and $m$ are charging slots and the number of UAVs that can be charged together in EV charging stations. The dash curves are the worst case of EVs, help to offload a large scale of UAVs to maximize the coverage probability and the solid lines are for the waiting time of EVs without infrastructure sharing.}
	\label{fig_Tw_EV}
\end{figure*}
\begin{figure*}[ht]
	\centering
	\includegraphics[width = 2\columnwidth]{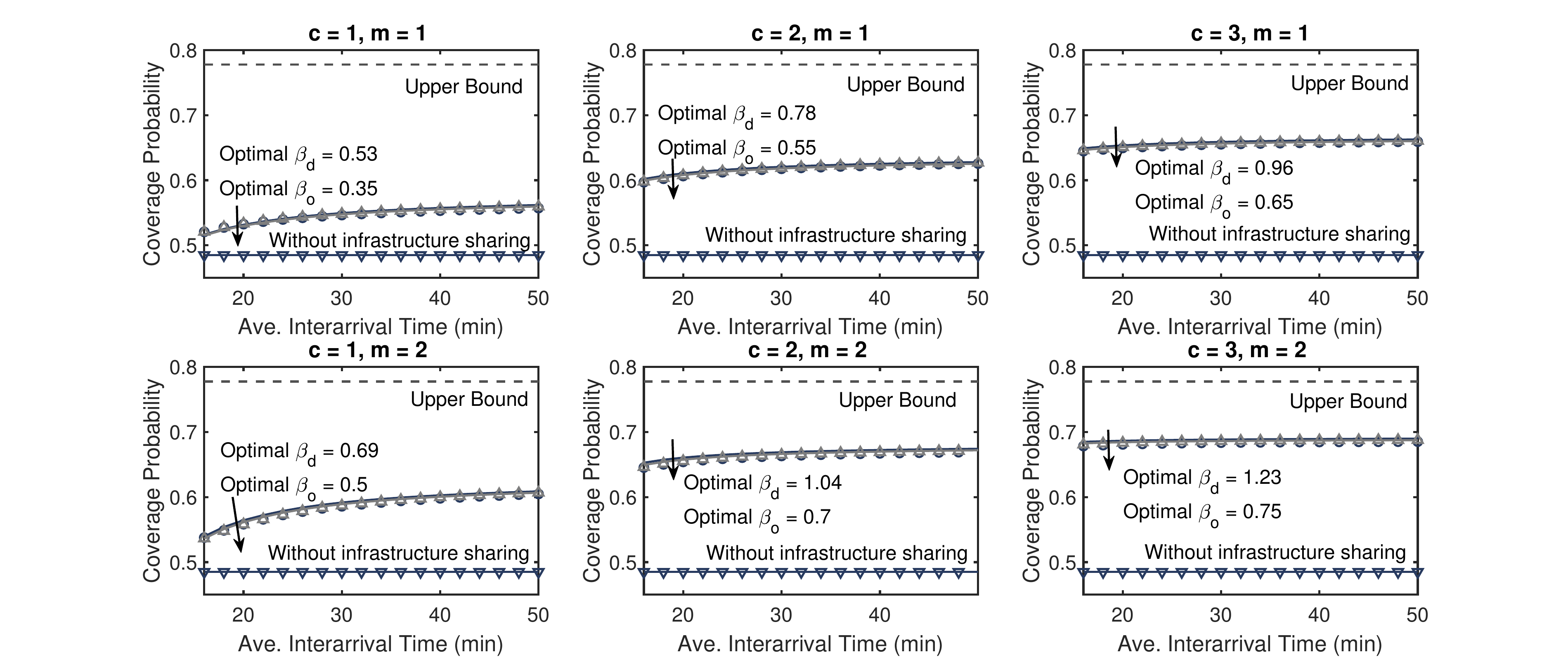}
	\caption{Analysis and simulation results of coverage probability of UAVs under different average interarrival time in the case of sharing and no sharing. $c$ and $m$ are charging slots and the number of UAVs that can be charged together in EV charging stations. The minima and maxima curves are the coverage probability of UAVs without infrastructure sharing and upper bound of given system (no waiting time and traveling time).}
	\label{fig_Pcov}
\end{figure*}
	
	In this section, we relate the system performance and the cost. While we show that infrastructure sharing can improve the coverage probability of UAVs, however, the operators also care about the fee they paid for the EV infrastructure sharing and the cost of extra dedicated charging stations. The same for EV operators, while they earn extra money from sharing, their own service quality decreases.
	
	As mentioned in Definition \ref{def_inf}, UAVs' operators pay for the EV infrastructure, we simply consider this fee is composed of voltage regulation and maintenance cost,
	\begin{align}
	&C_{\rm inf,d} = C_{\rm vol}+C_{\rm main}\nonumber\\
	&= \frac{365\times 24\times 60\times \mathbb{E}[N_{d,ev}]}{\mathbb{E}[T_{\rm ser}+T_{\rm ch}+T_{\rm tra}+T_{\rm w}]}\cdot B_{\rm max}\cdot c_{\rm vol}+c\cdot c_{\rm main},\label{eq_Cinfd_UAV}
	\end{align}
where $\frac{24\times 60}{\mathbb{E}[T_{\rm ser}+T_{\rm ch}+T_{\rm tra}+T_{\rm w}]}\times B_{\rm max}$ is the total energy UAV needed in a day, $c_{\rm vol}$ is the price EV operators are paid  and $c_{\rm main}$ is the maintenance cost per charger \cite{dominguez2019design}.

    EVs providers solve their optimization problem first: how many UAVs they can help to offload from UAV charging stations to maximize the profit from UAVs, given that the extra waiting time of EVs are tolerable. Then, UAVs operators solve their problem: based on the charging infrastructure EVs shared, how to minimize the cost while ensure an acceptable system performance, pay more for building their charging stations to achieve a better performance, or pay less for less dedicated charging stations and a lower system performance. 
    
 Observing that the objective functions (\ref{eq_Ce}) and (\ref{eq_Cu}) are functions of the electricity price, which is different at each hour of the day and varies from the region, and weights of objectives, it is difficult to obtain an exact value and say it provides the best performance. However, providing a general model to analyze this trade-off and considering the particularity of each realistic scenario is the most suitable solution to follow. Besides, the proposed system model and analysis can be easily extended to difference scenarios.  For instance, given the electricity price is different of the day, EVs' operators can change offloading UAV densities and UAVs' operators pay different amounts of fees for infrastructure sharing according to the electricity price, and then build their dedicated charging stations by jointly considering multiple offloading ratios. We do admit that such optimization highly depends on operators' decision and is parameter-based, our goal here is to capture the constraint of cost on system performance: deploying more charging stations do achieve better performances, but all these better performances are expected to cost more. 

	\section{Numerical Results}\label{sec_num}
	In this section we validate our analytical results with simulations and evaluate the improvement of system performance using infrastructure sharing. Unless stated otherwise, we use the simulation parameters as listed herein Table \ref{table_par}.
	
   	For the simulation of the considered setup, we apply Monte-Carlo simulations with a large number of iterations to ensure accuracy. In each iteration, we first generate $10^4$ exponential distributed random variables to simulate the arrival processes of EVs. To compute the number of UAVs associate with EV/UAV charging stations, we generate three independent PPPs and compute the association cell. The average waiting time of EVs and UAVs are obtained based on the two association policies and serving policies. We then generate another independent PPP realizations to model the locations of user cluster centers and TBSs and generate the locations of the reference user. Conditioned on the typical UAVs located at the origin, we derive the coverage probability. In all the figures we plot, markers are the simulation results and curves are the analysis results.

\begin{figure*}[ht]
	\centering
	\includegraphics[width = 1.5\columnwidth]{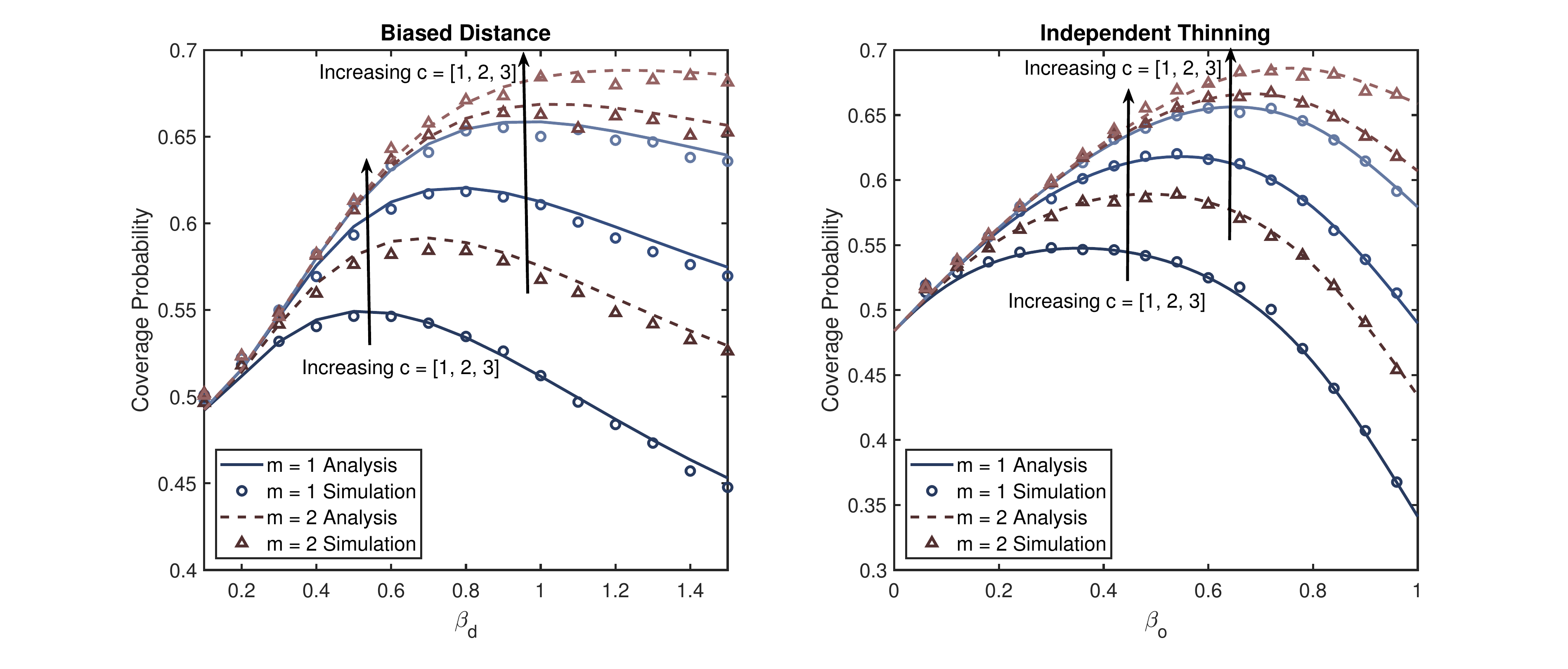}
	\caption{Coverage probability under different values of $\beta_{\{d,o\}}$. The dash curves are for $m = 2$ and the solid lines are for $m = 1$. Along with the arrow direction, $c$ increases.}
	\label{fig_OptBeta}
\end{figure*}

\begin{figure*}[ht]
	\centering
	\includegraphics[width = 2\columnwidth]{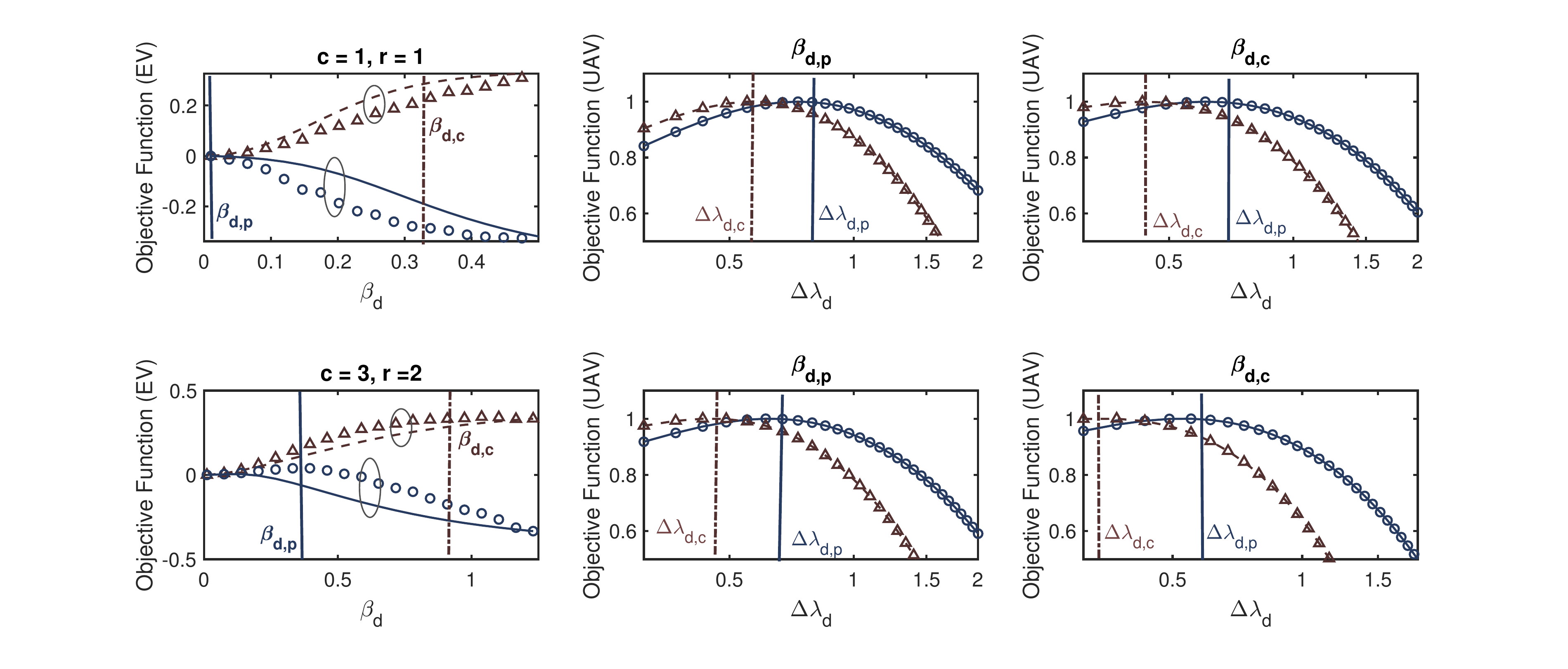}
	\caption{Results of objective functions. We consider two cases: operators care more about performance ($w_{\rm wait}>w_{\rm inf,ev}$ ,$w_{\rm cov}>w_{\rm c},w_{\rm inf,d}$) or more about profit, and corresponding decisions denoted by $\beta_{d,p},\Delta\lambda_{ d,p}$ and $\beta_{d,c},\Delta\lambda_{ d,c}$, respectively. All the performance-based decisions are plot in blue and sold lines with circle markers, and all the cost-based decisions are plot in red and dash lines with upward-pointing triangle markers.}
	\label{fig_OptInf}
\end{figure*}

In Fig. \ref{fig_Tw} and \ref{fig_Tw_EV}, we plot the waiting time of EVs and UAVs with/without infrastructure sharing: we increase $\beta_{\{d,o\}}$ from $0$, which means no infrastructure sharing, till the optimal values, which maximize the coverage probability, from the perspective of UAVs. Here we only plot the worst curve of the waiting time of EVs, under optimal values of $\beta_{\{d,o\}}$ (See Fig. \ref{fig_OptBeta} for details) since the waiting time does not change a lot. Interestingly, while the waiting time of UAVs drops dramatically, the waiting time of EVs does not increase a lot, as mentioned in Remark \ref{re_EV_first}, which is because that $T_{ w,ev}$ is only a function of $N_{d,ev}$ when $N_{d,ev}$ is a small value, while it stays constant when $N_{d,ev}$ is large, especially when the quality of EV charging station is high (large capacity and can charge multiple UAVs).

Fig. \ref{fig_Pcov} shows the benefit of infrastructure sharing for UAVs, as it improves the coverage probability and approaches the system upper bound where $T_{ w}$ and $T_{\rm ch}$ is 0. However, the coverage probability in the case of infrastructure sharing cannot reach the system upper bound, owing to a longer charging time in EV charging stations. That is, even though the waiting time is approximately 0 and traveling time is negligible, gap of availability probability (which is a time fraction denoted that UAV is available and can provide service) exists between using infrastructure sharing and deploying more dedicated charging stations. Hence it requires a trade-off between better performance and huge cost.

Fig. \ref{fig_OptBeta} shows the optimal values for both $\beta_d$ and $\beta_o$. As we offloading UAVs to EV charging stations, the waiting time in UAV charging stations decreases, hence the coverage probability increases. However, if we offload a large fraction of UAVs to EV charging stations, UAVs start to have a traffic and long waiting time in EV charging stations, then the coverage performance decreases. In both Fig. \ref{fig_Pcov} and Fig. \ref{fig_OptBeta}, we show that the system performance based on biased distance association policy is slightly better than independent thinning association policy, which is because of the traveling distance in the first association policy is shorter than than the second one.

Fig. \ref{fig_OptInf} shows the results of optimal infrastructure sharing strategy. While we find that the cost of installing extra dedicated charging stations is much higher than paying for the infrastructure sharing, we assume that UAVs' operators install extra charging stations (the density is denoted by $\Delta\lambda_{d}$) to improve its system performance based on the decision of $\beta_{d}$ (here we use $\beta_{d}$ to denote both $\beta_{d}$ and $\beta_{o}$, for simplification). With that being said, since we find infrastructure sharing fee is much lower than installing new charging stations, UAVs' operators deploy their dedicated charging stations based on the sharing strategy of EVs' operators.
Besides, the sharing fee paid for EVs' operators are not function of $\Delta\lambda_{d}$, it is a constant added up to the total cost. Hence, the optimization problem of EV, (\ref{eq_Ce}), is actually about a trade-off between performance and cost based on infrastructure sharing. If the operators care more about their services, less sharing ($\beta_{d,p}<\beta_{d,c}$) and more dedicated charging stations ($\Delta\lambda_{d,c}<\Delta\lambda_{d,p}$) deployed for EVs and UAVs respectively (blue and sold lines with circle markers).

As shown, when the quality of EV charging stations is poor (low capacity and cannot charge multiple UAVs), infrastructure sharing is not a good strategy since the waiting time of EVs increases dramatically. Hence, UAV operators need to build more dedicated charging stations to achieve an acceptable performance. When the quality of EV charging stations is high ($c = 3$, $r = 2$), EV charging stations can help to offload a larger scale of UAVs while still maintain the performance compared with low quality case.
 
\section{Conclusion}
 In this paper, we presented an optimization problem to analyze the possibility of sharing charging infrastructure in EV and UAV-involved networks. We first approximated waiting time for both EVs and UAVs in continuous time, and derived the availability probability of UAVs in a more accurate method, compared with existing literature. We then solve our optimization problem from the perspective of EVs' and UAVs' operators and based on weights of service quality and profit or cost. Our results show some interesting system insights: if the charging stations are high quality, charging infrastructure sharing can benefit both UAVs' and EVs' operators.
 
\appendix
\subsection{Waiting Time when $N_{d,ev}<mc(1+\frac{T_{\rm ser}}{T_{\rm ch,d,ev}})$}\label{app_Tw_less}
In this part, we provide the waiting time equations for $N_{d,ev}<mc(1+\frac{T_{\rm ser}}{T_{\rm ch,d,ev}})$. As mentioned in Remark \ref{re_Tw}, the probability of $N_{d,ev}<mc(1+\frac{T_{\rm ser}}{T_{\rm ch,d,ev}})$ happened is low and it is difficult to compute. Hence, we also present an approximation here. The following equations derived by conditioned on the number of EVs arrival during the serving time interval of the typical UAV, and then take the expectation.

In the case of $N_{d,ev}<mc(1+\frac{T_{\rm ser}}{T_{\rm ch,d,ev}})$, if $n \geq \frac{c(T_{\rm ser}-T_{\rm ch,d,ev})-T_{\rm ch,d,ev}\frac{N_{d,ev}}{m}}{\mathbb{E}[T_{\rm ch,ev}]}$,
\begin{align}
	T_{\rm w,d,ev|N_{d,ev}} &= \mathbb{E}_{n}\bigg[\frac{n\mathbb{E}[T_{\rm ch,ev}]}{c}+\bigg(\frac{\mathbb{E}[T_{\rm ch,ev}]\mu_e}{c}-1\bigg)\nonumber\\
		&\times\bigg(T_{\rm ser}-(\frac{N_{d,ev}}{mc}-1)T_{\rm ch,d,ev}\bigg)\bigg],  \nonumber
\end{align}
else,
\begin{align}
	&T_{\rm w,d,ev|N_{d,ev}} = \mathbb{E}_{n}\bigg[\sum_{x=0}^{\infty}(T_{\rm gap|n}\mu_e)^{x}\frac{\exp(-T_{\rm gap|n}\mu_e)}{x!}T_{w|x}\bigg],  \nonumber
\end{align}
where $n \sim Exp\bigg(T_{\rm w,d,ev|N_{d,ev}}+T_{\rm ch,d,ev}\frac{N_{d,ev}}{c}\bigg)$ and
\begin{align}
	&T_{\rm gap|n} = T_{\rm ser}-(\frac{N_{d,ev}}{mc}-1)T_{\rm ch,d}-\frac{\mathbb{E}[nT_{\rm ch,ev}]}{c},\nonumber\\
	&T_{w|x} =\int_{0}^{T_{\rm gap|n}}\int_{t_1}^{T_{\rm gap|n}}\cdots\int_{t_x}^{T_{\rm gap|n}}\frac{1}{c}f_{t_1,t_2\cdots,t_x}(t_1,t_2,\cdots)\nonumber\\
	&\quad \max\bigg(\max(\max(t_{1}^{'}+\mathbb{E}[T_{\rm ch,ev}],t_2)+\cdots,t_x)+\mathbb{E}[T_{\rm ch,ev}]\nonumber\\
	&\quad -T_{\rm gap|n},0\bigg){\rm d}t_x\cdots{\rm d}t_1,
\end{align}
when $x = 1,2,3$ the above equation $T_{w|x}$ is,
\begin{small}
	\begin{align}
		&\int_{0}^{T_{\rm gap|n}}\frac{1}{c}\frac{1}{T_{\rm gap|n}}\max(0,t_1^{'}+\mathbb{E}[T_{\rm ch,ev}]-T_{\rm gap|n}){\rm d}t_1,\\
		&\int_{0}^{T_{\rm gap|n}}\int_{t_1}^{T_{\rm gap|n}}\frac{1}{c}\frac{2}{T_{\rm gap|n}^2}\max(\max(t_{1}^{'}+\mathbb{E}[T_{\rm ch,ev}],t_2)\nonumber\\
		&\qquad\qquad\qquad+\mathbb{E}[T_{\rm ch,ev}]-T_{\rm gap|n},0){\rm d}t_2{\rm d}t_1,\\
		&\int_{0}^{T_{\rm gap|n}}\int_{t_1}^{T_{\rm gap|n}}\int_{t_2}^{T_{\rm gap|n}}\frac{1}{c}\frac{6}{T_{\rm gap|n}^3}\max(\max(\max(t_{1}^{'}\nonumber\\
		&+\mathbb{E}[T_{\rm ch,ev}],t_2)+\mathbb{E}[T_{\rm ch,ev}],t_3)+\mathbb{E}[T_{\rm ch,ev}]-T_{\rm gap|n},0){\rm d}t_3{\rm d}t_2{\rm d}t_1,
	\end{align}
\end{small}
for $x>3$, using the following lower bound approximation,
\begin{align}
\int_{0}^{T_{\rm gap|n}}\frac{x}{T_{\rm gap|n}}	&\frac{1}{c}(1-\frac{t_1}{T_{\rm gap|n}})^{x-1}\nonumber\\
	&\max(0,t_1^{'}+\frac{x}{c}\mathbb{E}[T_{\rm ch,ev}]-T_{\rm gap|n}){\rm d}t_1,
\end{align}
in which,
\begin{align}
	t_{1}^{'} &= (t_1<T)(T+t_1)+(t_1>T)t_1, \nonumber\\
	T &= \frac{n\mu_e\mathbb{E}^2[T_{\rm ch,ev}]}{c^2}.
\end{align}
\subsection{Proof of Theorem \ref{the_Pa_bias}}\label{app_Pa_bias}
	Conditioned on $N$ UAVs fill in the typical association cell, availability probability is given by,
	\begin{small}
\begin{align}
	&\mathbb{P}(\rm \mathcal{A}\mid N) \nonumber\\
	&= \mathcal{A}_{ev}(\beta)\mathbb{E}_{ Y_{ev}}\bigg[\frac{T_{\rm ser}(y)}{T_{\rm ser}(y)+T_{\rm ch,d,ev}+T_{\rm w,d,ev\mid N_{d,ev}}+2T_{\rm tra}(y)}\bigg]\nonumber\\
	&+ \mathcal{A}_{d}(\beta)\mathbb{E}_{ Y_{d}}\bigg[\frac{T_{\rm ser}(y)}{T_{\rm se}(y)+T_{\rm ch,d,d}+T_{\rm w,d,d\mid N_{d,d}}+2T_{\rm tra}(y)}\bigg]\nonumber\\
	&= \mathcal{A}_{d}(\beta)\mathbb{E}_{ Y_{d}}\bigg[\frac{vB_{\rm max}-2Y_{d}p_{m}}{vB_{\rm max}-2Y_{d}(p_{m}-p_{s})+vp_{s}(T_{\rm ch,d,d}+T_{\rm w,d,d\mid N_{d,d}})}\bigg]\nonumber\\
	&+ \mathcal{A}_{ev}(\beta)\nonumber\\
	&\quad\mathbb{E}_{ Y_{ev}}\bigg[\frac{vB_{\rm max}-2Y_{ev}p_{m}}{vB_{\rm max}-2Y_{ev}(p_{m}-p_{s})+vp_{s}(T_{\rm ch,d,ev}+T_{\rm w,d,ev\mid N_{d,ev}})}\bigg]\nonumber\\
	&= \int_{0}^{\frac{vB_{\rm max}}{2p_m}}\bigg[\frac{vB_{\rm max}-2yp_{m}}{vB_{\rm max}-2y(p_{m}-p_{s})+vp_{s}(T_{\rm ch,d,ev}+T_{\rm w,d\mid N_{d,ev}})}\bigg]\nonumber\\
	&\quad\times\bar{F}_{c,d}(\frac{y}{\beta})f_{c,ev}(y){\rm  d}y\nonumber\\
	&+ \int_{0}^{\frac{vB_{\rm max}}{2p_m}}\bigg[\frac{vB_{\rm max}-2yp_{m}}{vB_{\rm max}-2y(p_{m}-p_{s})+vp_{s}(T_{\rm ch,d,d}+T_{\rm w,d,d\mid N_{d,d}})}\bigg]\nonumber\\
	&\quad\times\bar{F}_{c,ev}(y\beta)f_{c,d}(y){\rm  d}y,	
\end{align}
\end{small}
then the unconditional availability probability is given by
\begin{align}
	P_{\rm a,bias} = &\sum_{n = 0}^{\infty}\mathbb{P}(\rm \mathcal{A}\mid N)\mathbb{P}(N = n), \nonumber
\end{align}
proof completes by taking the expectation over $N_{d,ev}$ and $N_{d,d}$, respectively.


\subsection{Proof of Lemma \ref{lem_user_ass}}\label{app_user_ass}
When the cluster UAV is not available, users associate with a nearby available UAV or the nearest TBSs, which provides the strongest average received power. The probability of associating with a nearby LoS UAV is,
	\begin{align}
	&\mathcal{A}_{\rm LoS}(r) = \mathcal{A}_{\rm LoS-NLoS}(r)\mathcal{A}_{\rm LoS-TBS}(r)\nonumber\\
	&= \mathbb{P}(\eta_l\rho_u r^{-\alpha_l}>\eta_n\rho_u R_{u_{n}^{'}}^{-\alpha_n})\mathbb{P}(\rho_u\eta_l  r^{-\alpha_l}>\rho_t R_{t}^{-\alpha_t})\nonumber\\
	&= \mathbb{P}\bigg(R_{u_{n}^{'}}>(\frac{\eta_n}{\eta_l})^{\frac{1}{\alpha_n}} r^{\frac{\alpha_l}{\alpha_n}}\bigg) \mathbb{P}\bigg(R_{t}>(\frac{\rho_t}{\rho_u\eta_l  })^{\frac{1}{\alpha_t}}r^{\frac{\alpha_l}{\alpha_t}}\bigg)\nonumber\\
	&= \exp\bigg(-2\pi\lambda_{u}^{'}\int_{0}^{\sqrt{d_n^2(r)-h^2}}zP_n(\sqrt{z^2+h^2}){\rm d}z\bigg)\nonumber\\
	&\quad\times\exp\bigg(-2\pi\lambda_{t}d_{lt}^2(r)\bigg),
\end{align}
where $d_{lt}(r)$ and $d_{n}(r)$ are defined in Lemma \ref{lem_user_ass}.
$\mathcal{A}_{\rm NLoS}(r)$ and $\mathcal{A}_{\rm TBS}(r)$ follow the same way, therefore omitted here.
	\bibliographystyle{IEEEtran}
	\bibliography{refrep5}
\end{document}

%% file: notation.tex
\def\nb0{{\mathbf{0}}}
\def\nb1{{\mathbf{1}}}

\newtheorem{lemma}{Lemma}

\newtheorem{definition}{Definition}

\newtheorem{theorem}{Theorem}

\newtheorem{remark}{Remark}

\def\erf{\operatorname{erf}}









%% file: summary.bbl
\begin{thebibliography}{10}
\providecommand{\url}[1]{#1}
\csname url@samestyle\endcsname
\providecommand{\newblock}{\relax}
\providecommand{\bibinfo}[2]{#2}
\providecommand{\BIBentrySTDinterwordspacing}{\spaceskip=0pt\relax}
\providecommand{\BIBentryALTinterwordstretchfactor}{4}
\providecommand{\BIBentryALTinterwordspacing}{\spaceskip=\fontdimen2\font plus
\BIBentryALTinterwordstretchfactor\fontdimen3\font minus
  \fontdimen4\font\relax}
\providecommand{\BIBforeignlanguage}[2]{{%
\expandafter\ifx\csname l@#1\endcsname\relax
\typeout{** WARNING: IEEEtran.bst: No hyphenation pattern has been}%
\typeout{** loaded for the language `#1'. Using the pattern for}%
\typeout{** the default language instead.}%
\else
\language=\csname l@#1\endcsname
\fi
#2}}
\providecommand{\BIBdecl}{\relax}
\BIBdecl

\bibitem{gan2012optimal}
L.~Gan, U.~Topcu, and S.~H. Low, ``Optimal decentralized protocol for electric
  vehicle charging,'' \emph{IEEE Transactions on Power Systems}, vol.~28,
  no.~2, pp. 940--951, 2012.

\bibitem{liu2021reservation}
S.~Liu, X.~Xia, Y.~Cao, Q.~Ni, X.~Zhang, and L.~Xu, ``Reservation-based {EV}
  charging recommendation concerning charging urgency policy,''
  \emph{Sustainable Cities and Society}, vol.~74, p. 103150, 2021.

\bibitem{sekander2018multi}
S.~Sekander, H.~Tabassum, and E.~Hossain, ``Multi-tier drone architecture for
  {5G/B5G} cellular networks: Challenges, trends, and prospects,'' \emph{IEEE
  Communications Magazine}, vol.~56, no.~3, pp. 96--103, 2018.

\bibitem{mozaffari2019tutorial}
M.~Mozaffari, W.~Saad, M.~Bennis, Y.-H. Nam, and M.~Debbah, ``A tutorial on
  {UAV}s for wireless networks: Applications, challenges, and open problems,''
  \emph{IEEE communications surveys \& tutorials}, vol.~21, no.~3, pp.
  2334--2360, 2019.

\bibitem{li2018uav}
B.~Li, Z.~Fei, and Y.~Zhang, ``{UAV} communications for {5G} and beyond: Recent
  advances and future trends,'' \emph{IEEE Internet of Things Journal}, vol.~6,
  no.~2, pp. 2241--2263, 2018.

\bibitem{zeng2016wireless}
Y.~Zeng, R.~Zhang, and T.~J. Lim, ``Wireless communications with unmanned
  aerial vehicles: Opportunities and challenges,'' \emph{IEEE Communications
  Magazine}, vol.~54, no.~5, pp. 36--42, 2016.

\bibitem{9773146}
Y.~Qin, M.~A. Kishk, and M.-S. Alouini, ``Drone charging stations deployment in
  rural areas for better wireless coverage: Challenges and solutions,''
  \emph{IEEE Internet of Things Magazine}, vol.~5, no.~1, pp. 148--153, 2022.

\bibitem{liao2020learning}
H.~Liao, Z.~Zhou, W.~Kong, Y.~Chen, X.~Wang, Z.~Wang, and S.~Al~Otaibi,
  ``Learning-based intent-aware task offloading for air-ground integrated
  vehicular edge computing,'' \emph{IEEE Transactions on Intelligent
  Transportation Systems}, 2020.

\bibitem{cheng2019space}
N.~Cheng, F.~Lyu, W.~Quan, C.~Zhou, H.~He, W.~Shi, and X.~Shen,
  ``Space/aerial-assisted computing offloading for {IoT} applications: A
  learning-based approach,'' \emph{IEEE Journal on Selected Areas in
  Communications}, vol.~37, no.~5, pp. 1117--1129, 2019.

\bibitem{matracia2021topological}
M.~Matracia, M.~A. Kishk, and M.-S. Alouini, ``On the topological aspects of
  {UAV}-assisted post-disaster wireless communication networks,'' \emph{IEEE
  Communications Magazine}, vol.~59, no.~11, pp. 59--64, 2021.

\bibitem{9205314}
M.~Kishk, A.~Bader, and M.-S. Alouini, ``Aerial base station deployment in 6{G}
  cellular networks using tethered drones: The mobility and endurance
  tradeoff,'' \emph{IEEE Vehicular Technology Magazine}, vol.~15, no.~4, pp.
  103--111, 2020.

\bibitem{8673613}
Q.~Cui, Y.~Weng, and C.-W. Tan, ``Electric vehicle charging station placement
  method for urban areas,'' \emph{IEEE Transactions on Smart Grid}, vol.~10,
  no.~6, pp. 6552--6565, 2019.

\bibitem{SHAUKAT20181329}
\BIBentryALTinterwordspacing
N.~Shaukat, B.~Khan, S.~Ali, C.~Mehmood, J.~Khan, U.~Farid, M.~Majid, S.~Anwar,
  M.~Jawad, and Z.~Ullah, ``A survey on electric vehicle transportation within
  smart grid system,'' \emph{Renewable and Sustainable Energy Reviews},
  vol.~81, pp. 1329--1349, 2018. [Online]. Available:
  \url{https://www.sciencedirect.com/science/article/pii/S1364032117307190}
\BIBentrySTDinterwordspacing

\bibitem{sweda2011agent}
T.~Sweda and D.~Klabjan, ``An agent-based decision support system for electric
  vehicle charging infrastructure deployment,'' in \emph{2011 IEEE Vehicle
  Power and Propulsion Conference}.\hskip 1em plus 0.5em minus 0.4em\relax
  IEEE, 2011, pp. 1--5.

\bibitem{9681355}
S.~Hussain, Y.-S. Kim, S.~Thakur, and J.~G. Breslin, ``Optimization of waiting
  time for electric vehicles using a fuzzy inference system,'' \emph{IEEE
  Transactions on Intelligent Transportation Systems}, pp. 1--12, 2022.

\bibitem{oda2018mitigation}
T.~Oda, M.~Aziz, T.~Mitani, Y.~Watanabe, and T.~Kashiwagi, ``Mitigation of
  congestion related to quick charging of electric vehicles based on waiting
  time and cost--benefit analyses: A japanese case study,'' \emph{Sustainable
  cities and society}, vol.~36, pp. 99--106, 2018.

\bibitem{sathaye2013approach}
N.~Sathaye and S.~Kelley, ``An approach for the optimal planning of electric
  vehicle infrastructure for highway corridors,'' \emph{Transportation Research
  Part E: Logistics and Transportation Review}, vol.~59, pp. 15--33, 2013.

\bibitem{6486056}
S.~Chen and L.~Tong, ``i{EMS} for large scale charging of electric vehicles:
  Architecture and optimal online scheduling,'' in \emph{2012 IEEE Third
  International Conference on Smart Grid Communications (SmartGridComm)}, 2012,
  pp. 629--634.

\bibitem{6557596}
J.~J. Yu, V.~O. Li, and A.~Y. Lam, ``Optimal {V2G} scheduling of electric
  vehicles and unit commitment using chemical reaction optimization,'' in
  \emph{2013 IEEE Congress on Evolutionary Computation}, 2013, pp. 392--399.

\bibitem{6486024}
A.~Y. Lam, K.-C. Leung, and V.~O. Li, ``Capacity management of vehicle-to-grid
  system for power regulation services,'' in \emph{2012 IEEE Third
  International Conference on Smart Grid Communications (SmartGridComm)}, 2012,
  pp. 442--447.

\bibitem{7733098}
H.~ElSawy, A.~Sultan-Salem, M.-S. Alouini, and M.~Z. Win, ``Modeling and
  analysis of cellular networks using stochastic geometry: A tutorial,''
  \emph{IEEE Communications Surveys Tutorials}, vol.~19, no.~1, pp. 167--203,
  2017.

\bibitem{6524460}
H.~ElSawy, E.~Hossain, and M.~Haenggi, ``Stochastic geometry for modeling,
  analysis, and design of multi-tier and cognitive cellular wireless networks:
  A survey,'' \emph{IEEE Communications Surveys Tutorials}, vol.~15, no.~3, pp.
  996--1019, 2013.

\bibitem{9153823}
Y.~Qin, M.~A. Kishk, and M.-S. Alouini, ``Performance evaluation of
  {UAV}-enabled cellular networks with battery-limited drones,'' \emph{IEEE
  Communications Letters}, vol.~24, no.~12, pp. 2664--2668, 2020.

\bibitem{9444343}
------, ``On the influence of charging stations spatial distribution on aerial
  wireless networks,'' \emph{IEEE Transactions on Green Communications and
  Networking}, vol.~5, no.~3, pp. 1395--1409, 2021.

\bibitem{8866716}
M.-A. Lahmeri, M.~A. Kishk, and M.-S. Alouini, ``Stochastic geometry-based
  analysis of airborne base stations with laser-powered {UAV}s,'' \emph{IEEE
  Communications Letters}, vol.~24, no.~1, pp. 173--177, 2020.

\bibitem{sekander2020statistical}
S.~Sekander, H.~Tabassum, and E.~Hossain, ``Statistical performance modeling of
  solar and wind-powered {UAV} communications,'' \emph{IEEE Transactions on
  Mobile Computing}, 2020.

\bibitem{kishk20203}
M.~A. Kishk, A.~Bader, and M.-S. Alouini, ``On the {3-D} placement of airborne
  base stations using tethered {UAV}s,'' \emph{IEEE Transactions on
  Communications}, vol.~68, no.~8, pp. 5202--5215, 2020.

\bibitem{lou2021green}
Z.~Lou, A.~Elzanaty, and M.-S. Alouini, ``Green tethered {UAV}s for {EMF}-aware
  cellular networks,'' \emph{IEEE Transactions on Green Communications and
  Networking}, vol.~5, no.~4, pp. 1697--1711, 2021.

\bibitem{meddour2011role}
D.-E. Meddour, T.~Rasheed, and Y.~Gourhant, ``On the role of infrastructure
  sharing for mobile network operators in emerging markets,'' \emph{Computer
  networks}, vol.~55, no.~7, pp. 1576--1591, 2011.

\bibitem{sanguanpuak2018infrastructure}
T.~Sanguanpuak, S.~Guruacharya, E.~Hossain, N.~Rajatheva, and M.~Latva-aho,
  ``Infrastructure sharing for mobile network operators: Analysis of trade-offs
  and market,'' \emph{IEEE Transactions on Mobile Computing}, vol.~17, no.~12,
  pp. 2804--2817, 2018.

\bibitem{bousia2015game}
A.~Bousia, E.~Kartsakli, A.~Antonopoulos, L.~Alonso, and C.~Verikoukis,
  ``Game-theoretic infrastructure sharing in multioperator cellular networks,''
  \emph{IEEE Transactions on Vehicular Technology}, vol.~65, no.~5, pp.
  3326--3341, 2015.

\bibitem{8315130}
R.~Jurdi, A.~K. Gupta, J.~G. Andrews, and R.~W. Heath, ``Modeling
  infrastructure sharing in mm{Wave} networks with shared spectrum licenses,''
  \emph{IEEE Transactions on Cognitive Communications and Networking}, vol.~4,
  no.~2, pp. 328--343, 2018.

\bibitem{fund2016spectrum}
F.~Fund, S.~Shahsavari, S.~S. Panwar, E.~Erkip, and S.~Rangan, ``Spectrum and
  infrastructure sharing in millimeter wave cellular networks: An economic
  perspective,'' \emph{arXiv preprint arXiv:1605.04602}, 2016.

\bibitem{dominguez2019design}
J.~Domnguez-Navarro, R.~Dufo-Lpez, J.~Yusta-Loyo, J.~Artal-Sevil, and
  J.~Bernal-Agust{\'\i}n, ``Design of an electric vehicle fast-charging station
  with integration of renewable energy and storage systems,''
  \emph{International Journal of Electrical Power \& Energy Systems}, vol. 105,
  pp. 46--58, 2019.

\bibitem{zeng2019energy}
Y.~Zeng, J.~Xu, and R.~Zhang, ``Energy minimization for wireless communication
  with rotary-wing {UAV},'' \emph{IEEE Transactions on Wireless
  Communications}, vol.~18, no.~4, pp. 2329--2345, 2019.

\bibitem{al2014optimal}
A.~Al-Hourani, S.~Kandeepan, and S.~Lardner, ``Optimal {LAP} altitude for
  maximum coverage,'' \emph{IEEE Wireless Communications Letters}, vol.~3,
  no.~6, pp. 569--572, 2014.

\bibitem{singh2013offloading}
S.~Singh, H.~S. Dhillon, and J.~G. Andrews, ``Offloading in heterogeneous
  networks: Modeling, analysis, and design insights,'' \emph{IEEE transactions
  on wireless communications}, vol.~12, no.~5, pp. 2484--2497, 2013.

\bibitem{alzer1997some}
H.~Alzer, ``On some inequalities for the incomplete {G}amma function,''
  \emph{Mathematics of Computation}, vol.~66, no. 218, pp. 771--778, 1997.

\bibitem{alzenad2019coverage}
M.~Alzenad and H.~Yanikomeroglu, ``Coverage and rate analysis for vertical
  heterogeneous networks ({VHetNet}s),'' \emph{IEEE Transactions on Wireless
  Communications}, vol.~18, no.~12, pp. 5643--5657, 2019.

\bibitem{bai2014coverage}
T.~Bai and R.~W. Heath, ``Coverage and rate analysis for millimeter-wave
  cellular networks,'' \emph{IEEE Transactions on Wireless Communications},
  vol.~14, no.~2, pp. 1100--1114, 2014.

\end{thebibliography}
